\documentclass[fleqn,10pt,twocolumn]{wlscirep}
\usepackage{placeins}	
\usepackage{textcomp}
\usepackage{hyperref}
\usepackage{float}
\usepackage{parskip}
\usepackage{siunitx}
\usepackage{enumerate}
\usepackage{mathtools}
\usepackage{stmaryrd}

\usepackage{enumitem}
\usepackage{bibunits}
\defaultbibliographystyle{naturemag-doi}
\defaultbibliography{ms_RJ}  

\usepackage{float}
\usepackage{textcomp}
\usepackage{gensymb}

\newcommand{\dd}[0]{\mathrm{d}}

\newcommand{\TT}[0]{\boldsymbol{T}}

\newcommand{\rr}[0]{\mathbf{r}}

\newcommand{\kB}[0]{k_{\mathrm{B}}}

\newcommand{\nn}[0]{\hat{\mathbf{n}}}
\newcommand{\uu}[0]{\hat{\mathbf{u}}}

\newcommand{\rotop}[0]{\mathbfcal{R}}

\newcommand\trick[1]{}

\usepackage{bm}
\DeclareMathAlphabet\mathbfcal{OMS}{cmsy}{b}{n}

\usepackage{hyperref}
\definecolor{mygreen}{rgb}{0.01, 0.5, 0.01}
\definecolor{myred}{rgb}{0.8, 0.4745098039215686, 0.6549019607843137}
\hypersetup{
    colorlinks=true,
    linkcolor=mygreen,
    citecolor=mygreen
}

\RequirePackage[colorlinks=true, allcolors=blue]{hyperref}

\newcommand{\rj}[1]{\textcolor{black}{#1}}
\newcommand{\rjj}[1]{\textcolor{black}{#1}}

\title{
Controlling microalgae populations by phototactic memory
}

\author[a,b,c,d, *]{Gianni Jacucci}
\author[e,f]{Davide Breoni}
\author[g]{Pierre Illien}
\author[e,f]{Luca Tubiana}
\author[a]{Jean-François Allemand}
\author[h]{Sylvain Gigan} 
\author[a, *]{Raphaël Jeanneret}

\affil[a]{Laboratoire de Physique de l’Ecole Normale Supérieure, ENS, Université PSL, CNRS, Sorbonne Université, Université Paris Cité, 75005 Paris, France}

\affil[b]{Dipartimento di Fisica, Università della Calabria, 87036 Arcavacata di Rende (CS), Italy}

\affil[c]{Istituto di Nanotecnologia, Consiglio Nazionale delle Ricerche (CNR-Nanotec), 87036 Arcavacata di Rende (CS), Italy}

\affil[d]{INFN, gruppo collegato di Cosenza, 87036 Arcavacata di Rende (CS), Italy}

\affil[e]{\textcolor{black}{Department of Physics, Università di Trento, Via Sommarive 14, I-38123 Trento, Italy }}

\affil[f]{\textcolor{black}{INFN-TIFPA, Trento Institute for Fundamental Physics and Applications, I-38123 Trento, Italy}}

\affil[g]{Sorbonne Université, CNRS, Physical Chemistry of Electrolytes and Interfacial Nanosystems (PHENIX), 4 place Jussieu, Paris, France}

\affil[h]{Laboratoire Kastler Brossel, Sorbonne Université, Ecole Normale Supérieure-Paris Sciences et Lettres (PSL) Research University, Centre National de la Recherche Scientifique (CNRS) UMR 8552, Collège de France, 24 rue Lhomond, 75005 Paris, France}

\affil[*]{To whom correspondence should be addressed. E-mail: 

giovanni.iacucci@phys.ens.fr; raphael.jeanneret@phys.ens.fr}

\begin{abstract}
Motile microorganisms navigate environments that vary in space and time by coupling sensory inputs to internal physiological states and active motion. In photosynthetic microorganisms, such coupling can be mediated by phototaxis, yet how temporal sensing combines with spatial variations of the environment to shape population organisation remains poorly understood. To probe this coupling, we expose populations of the phototactic microalga \textit{Chlamydomonas reinhardtii} to structured light landscapes. Under ring-shaped illumination, cells robustly accumulate at the center of the illuminated region, rather than at the light-intensity maximum. Single-cell tracking shows that cells steer relative to local intensity gradients and undergo a non-trivial phototactic response, where, on average, cells outside the ring perform positive phototaxis while cells inside perform negative phototaxis.
Population experiments, numerical simulations and analytical modelling show that this trapping cannot be reproduced by an instantaneous switching rule, but requires a finite phototactic memory acting on the transition between positive and negative phototaxis. This internal timescale couples to the geometry of the light field and to the persistence of active motion, enabling controllable spatial localisation without physical confinement. Our results generically show how memory in a single-cell sensory response can control population-level organisation in a living active system.
\end{abstract}

\begin{document}
\maketitle

\begin{bibunit}[naturemag-doi]

\section*{Introduction}
The ability to sense and respond to environmental cues is fundamental to the survival of living systems, enabling microorganisms and cells to locate favourable conditions and avoid adverse stimuli. To achieve this, they adapt their motility through a variety of taxis mechanisms, like chemotaxis \cite{bergChemotaxisEscherichiaColi1972}, gravitaxis \cite{robertsGravitaxisMotileMicroorganisms2002}, rheotaxis \cite{marcosBacterialRheotaxis2012} or phototaxis \cite{Foster1980}.
In particular, the latter plays a key role in microalgal ecology, enabling cells to optimise photosynthesis while minimising light-induced stress \cite{eberhardDynamicsPhotosynthesis2008, ericksonLightStressPhotoprotection2015}.

When compared to chemotaxis, which has been extensively studied \cite{Adler1966, Berg1975, berg2004coli, Wadhams2004}, the mechanisms underlying phototaxis \rjj{remain} elusive. 
Most studies aiming at investigating how photosensitive cells navigate in response to light focused on the model organism \textit{Chlamydomonas reinhardtii} \cite{alma990094274380107871, jeanneretBriefIntroductionModel2016}, a unicellular, motile organism.
Phototaxis in \textit{C.~reinhardtii} is mediated by a specialised organelle known as the eyespot—a photosensitive structure that allows cells to detect the direction of illumination \cite{kateriyaVisionSingleCelledAlgae2004}. As the cell rotates along its body axis \cite{Foster1980}, the changing orientation of the eyespot relative to the light source leads to a modulation of the detected signal \cite{yoshimuraSensitivityChlamydomonasPhotoreceptor2001}. 
This information enables the organism to steer towards or away from the light \rjj{\cite{schallerHowChlamydomonasKeeps1997,Goldstein2023, Goldstein2025}}. The resulting directional response is \rj{relatively} well characterised at the single-cell level and provides a foundational framework for understanding phototactic navigation. 
In addition, the eyespot apparatus has been extensively studied at the molecular level, from its biosynthesis and ultrastructure \cite{kreimer_green_2009}, to photocurrents and membrane depolarization triggered by Channelrhodopsins excitation \cite{hegemann_algal_2008}, to their ion selectivity \cite{baidukova_gating_2022} and phosphorylation states involved in phototactic adaptation \cite{bohm_channelrhodopsin-1_2019}, to light-dependent regulation of their abundance \cite{wolfram_multifactorial_2023,wolfram_insights_2024}. Despite this progress, a comprehensive mechanistic connection between these molecular details and cell behaviour is still lacking.

While phototactic navigation in simple fields is relatively well characterised \cite{CHOUDHARY20191508,Goldstein2023,Laroussi2024,Tsang2025, Goldstein2025}, natural environments seldom present uniform, purely directional light fields. How phototactic microorganisms navigate such complex cues remains poorly understood. In particular, how cells sense and respond to light intensity gradients—rather than directional cues—is unclear \cite{Arrieta2017,Brunet2022}.
Although gradient sensing and history-dependent responses have been extensively studied in the context of chemotaxis \cite{friedrichChemotaxisSpermCells2007, celaniBacterialStrategiesChemotaxis2010, kromerDecisionMakingImproves2018, gosztolaiCellularMemoryEnhances2020, karmakarCellularMemoryEukaryotic2021, colinMultipleFunctionsFlagellar2021}, their relevance to phototactic behaviour has been overlooked.
These questions are especially pertinent when
considering how individual sensing strategies determine the migration patterns of microalgae populations under non-uniform illumination.

\begin{figure}
    \centering
    \includegraphics[width=\columnwidth]{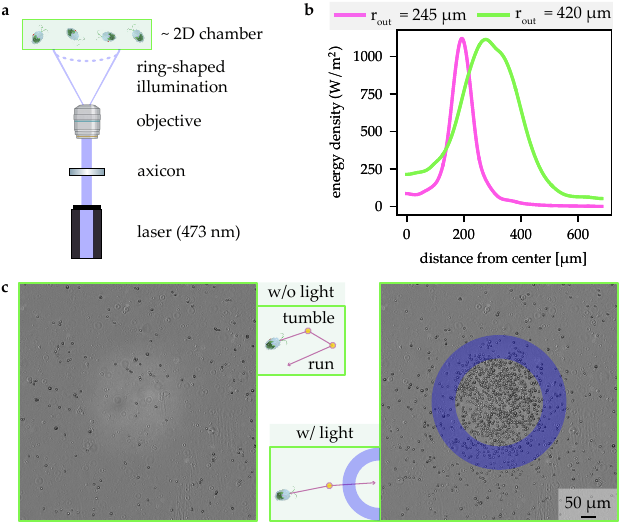}
    \caption{\textbf{Light-controlled spatial organisation of microalgae.}
    \textbf{a)} Schematic of the experimental setup. \textit{Chlamydomonas reinhardtii} cells ($\sim$\SI{10}{\micro\meter}) are confined in a quasi-two-dimensional microfluidic chamber ($\sim$\SI{20}{\micro\meter} thick) and illuminated with a ring-shaped blue light pattern generated by an axicon lens and a microscope objective.
    \textbf{b)} Energy density profiles of two ring illuminations with same inner radius ($r_{in} = $\SI{150}{\micro\meter}) and peak intensity, but different outer radii ($r_{out}$): \SI{245}{\micro\meter} (magenta) and \SI{420}{\micro\meter} (green).
    \textbf{c)} Bright-field images of the microalgae \rjj{suspension} without illumination (left panel) and after 1 hour of exposure to the ring-shaped pattern (right panel). In the absence of light, cells are uniformly distributed and display run-and-tumble motility. Upon illumination, they actively migrate towards the ring and accumulate within it, demonstrating light-guided spatial organisation. Source data are provided as a Source Data file.
    }
    \label{fig1}
\end{figure}
Here, we use spatially structured light to investigate phototactic navigation and control the spatial organisation of \textit{C.~reinhardtii} populations. By analysing single-cell trajectories, we show that cells actively steer along light gradients, and that their \rj{phototactic behaviour} is shaped by a short-time memory of recent illumination \cite{Laroussi2024}. This mechanism enables targeted accumulation of microalgae in predefined regions as a consequence of the subtle interplay between memory effects, swimming persistence and light field geometry. This provides a strategy to manipulate the spatial distribution of phototactic populations through external optical cues, and opens the way to finely control other types of tactic agents.
These findings highlight how structured environments can be harnessed to both reveal and control navigation strategies in active systems, offering a route to programmable behaviour in biological and synthetic microswimmers.

\section*{Results}
\subsection*{Light-controlled microalgae accumulation}
To study phototaxis in \textit{C.~reinhardtii} (\SI{\sim 10}{\micro\metre} in size), we confined populations in a quasi-two-dimensional microfluidic chamber (\SI{\sim 20}{\micro\metre} in height, \SI{\sim \rj{5}}{\milli\metre} in width), which ensures that cells swim in a well-defined plane while experiencing a controlled light environment (\autoref{fig1}\textcolor{mygreen}{a}). To structure this environment, a laser beam (\SI{473}{\nano\meter}) was shaped into a ring using an axicon lens (\autoref{fig1}\textcolor{mygreen}{a}), creating a blue-light pattern on the sample plane with tunable ring parameters (e.g. inner and outer radii), a central intensity well, and broad illumination tails (\autoref{fig1}\textcolor{mygreen}{b}, \textcolor{mygreen}{Figure S1}, and Methods).
The illumination is applied perpendicular to the plane of motion of the microalgae, 
\rj{preventing cells from aligning with the light propagation direction. 
This configuration allows us to investigate the microscopic response of the algae to light intensity gradients.}

In the absence of light, cells were uniformly distributed across the chamber, performing run-and-tumble motion characterised by alternating phases of straight swimming and random reorientation (\autoref{fig1}\textcolor{mygreen}{c}, left panel). 
In general, \textit{C.~reinhardtii} exhibits both positive and negative phototaxis, with the sign of the response depending on the local light intensity relative to a characteristic threshold \cite{Brunet2022,Arrieta2017}.
Upon switching on the illumination, the number of cells within the field of view increased markedly, as microalgae migrated in from surrounding regions towards the illuminated zone. Following this influx, cells accumulated inside the ring, beyond the location of maximum light intensity \rj{(\autoref{fig1}\textcolor{mygreen}{c}, right panel)}—an observation that points to a transition from positive to negative phototaxis as they cross the \rj{ring of maximal intensity.}

\begin{figure*}
    \centering
    \includegraphics[width=\linewidth]{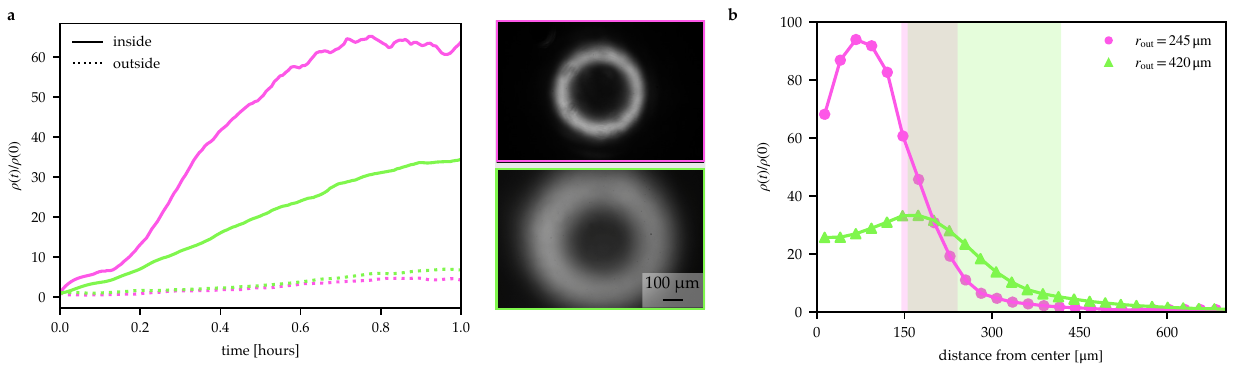}
    \caption{\textbf{Dynamics of microalgae accumulation under ring-shaped illumination.}
    Comparison of the accumulation dynamics under illuminations with the same inner radius (\SI{150}{\micro\meter}) but different outer radii, \SI{245}{\micro\meter} and \SI{420}{\micro\meter}, shown in pink and green, respectively.
    \textbf{a)} Temporal evolution of the normalised cell number density inside (solid lines) and outside (dotted lines) the illuminated region. In both cases, the density inside the ring rises sharply following illumination. The density outside the ring also increases, indicating active migration of cells toward the illuminated region from areas beyond the field of view. Accumulation is faster and more pronounced in the thinner ring.
    \textbf{b)} Spatial density profiles measured one hour into the experiment. The shaded regions indicate the spatial extent of the illumination, corresponding to the central peak and extending across the full width at half maximum of the intensity profile. As the ring thickness increases, the density distribution transitions from a pronounced peak inside the ring to a broader, plateau-like profile. Source data are provided as a Source Data file.}
    \label{fig2}
\end{figure*}

Having observed that microalgae migrated towards and accumulated inside the ring, we next quantified the dynamics of this accumulation, focusing on the role of the ring geometry (\autoref{fig2}\textcolor{mygreen}{a}). We monitored the temporal evolution of the normalised cell number density inside and outside the illuminated area for rings with identical inner radii (\SI{150}{\micro\metre}) and \rj{peak intensity} but differing outer radii (\SI{245}{\micro\metre} and \SI{420}{\micro\metre}, corresponding to a thin and a thick ring, respectively, \rj{\autoref{fig1}\textcolor{mygreen}{b}}). 
In both cases, the density inside the ring peaked within the first hour.
At the same time, the density outside the ring increased steadily, indicating long-range migration into the illuminated area. 
Comparing the two ring geometries, we \rj{found} that the thin ring leads to a faster accumulation rate and a higher cell number density.
Density profiles measured after one hour (\autoref{fig2}\textcolor{mygreen}{b}) reveal the effect of ring thickness on the spatial distribution of the microalgae. Increasing the ring thickness leads to a broader, plateau-like spatial density profile, in contrast to the sharply peaked distribution observed for narrower illumination. In both cases, cells accumulate within the ring, with peak density occurring away from the location of maximum light intensity.

\rjj{\subsection*{Gradient sensing and alignment}}
To investigate the mechanism underlying the observed accumulation, we analysed the trajectories of individual cells and characterised their phototactic response as a function of distance from the centre of the illumination. Given that cells exhibit some type of run-and-tumble motility in the dark \cite{polin_chlamydomonas_2009}, alternating straight swimming runs with rapid reorientation events, we specifically examined how the swimming direction and run duration correlated with the local light gradient.
Defining tumbles as in \cite{polin_chlamydomonas_2009}, we first measured the run duration as a function of the angle between the swimming direction and the gradient ($\theta$ in \autoref{fig3}\textcolor{mygreen}{a}). For cells located outside the ring (distances from center $d$ greater than \SI{300}{\micro\metre}), the run duration is maximised when the microalgae movement is aligned along the gradient ($\theta = 0$), confirming that cells spend more time swimming towards regions of higher light intensity. This anisotropy indicates active steering along the gradient. In chemotaxis, a similar modulation of run durations is used by motile bacteria, which bias their motion by extending or shortening runs depending on whether they move up or down chemical gradients \cite{Berg1975}. 
While $\textit{C.~reinhardtii}$ is known to steer using \rj{light direction} detected via its eyespot, our observations demonstrate \rj{how} it can adjust its phototactic response by exploiting local changes in light intensity \cite{Brunet2022, Arrieta2017}.

The preferential alignment observed in the run durations is also evident in the probability distribution of \rj{instantaneous} swimming angles relative to the gradient ($P(\theta)$ in \autoref{fig3}\textcolor{mygreen}{b}). Outside the ring, $P(\theta)$ is peaked around $\theta = 0$, consistent with positive phototaxis. This peak reflects the longer run times along the gradient (\autoref{fig3}\textcolor{mygreen}{a}). As cells \rj{cross} the ring, the distribution shifts, peaking near $\theta = \pi$ inside the ring and revealing a switch to negative phototaxis. This transition occurs in the vicinity of the maximum light intensity and provides a mechanistic explanation for the accumulation of cells inside the ring.

\subsection*{Memory-driven localisation}
\begin{figure*}
    \centering
    \includegraphics[width=\linewidth]{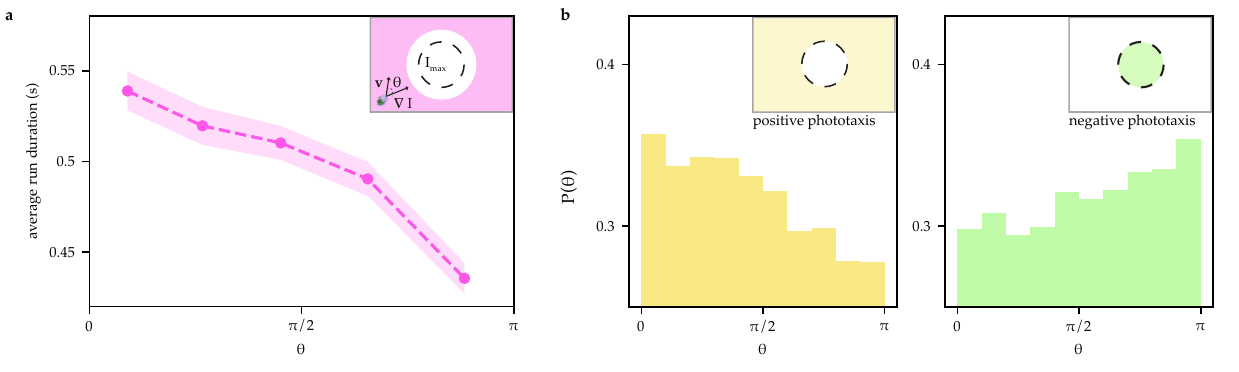}
    \caption{
    \textbf{Statistics of single-cell phototaxis.} 
    \textbf{a)} Microalgae located outside the ring-shaped illumination (distance from centre $d>$ \SI{300}{\micro\meter}) exhibit \rjj{biased} swimming towards the high-intensity region. The run duration is maximised when cells are aligned with the light gradient ($\theta = 0$), indicating active steering towards the ring. Pink shading represents the SE. 
    \textbf{b)} Probability distribution of the swimming angle relative to the local light gradient (P$(\theta)$) as a function of the distance from the ring centre. Two distinct phototactic regimes emerge: outside the ring, cells preferentially swim along the light gradient ($\theta = 0$), indicating positive phototaxis; inside the ring, the response switches to negative phototaxis, with a preference for swimming away from the gradient ($\theta = \pi$). 
    Data were obtained from averaging at least 100 trajectories, each lasting 2 minutes, and an illumination with $r_{in} = \SI{150}{\micro\meter}$ and $r_{out} = \SI{250}{\micro\meter}$. Source data are provided as a Source Data file.
    }
    \label{fig3}
\end{figure*}

\begin{figure*}
    \centering
    \includegraphics[width=\linewidth]{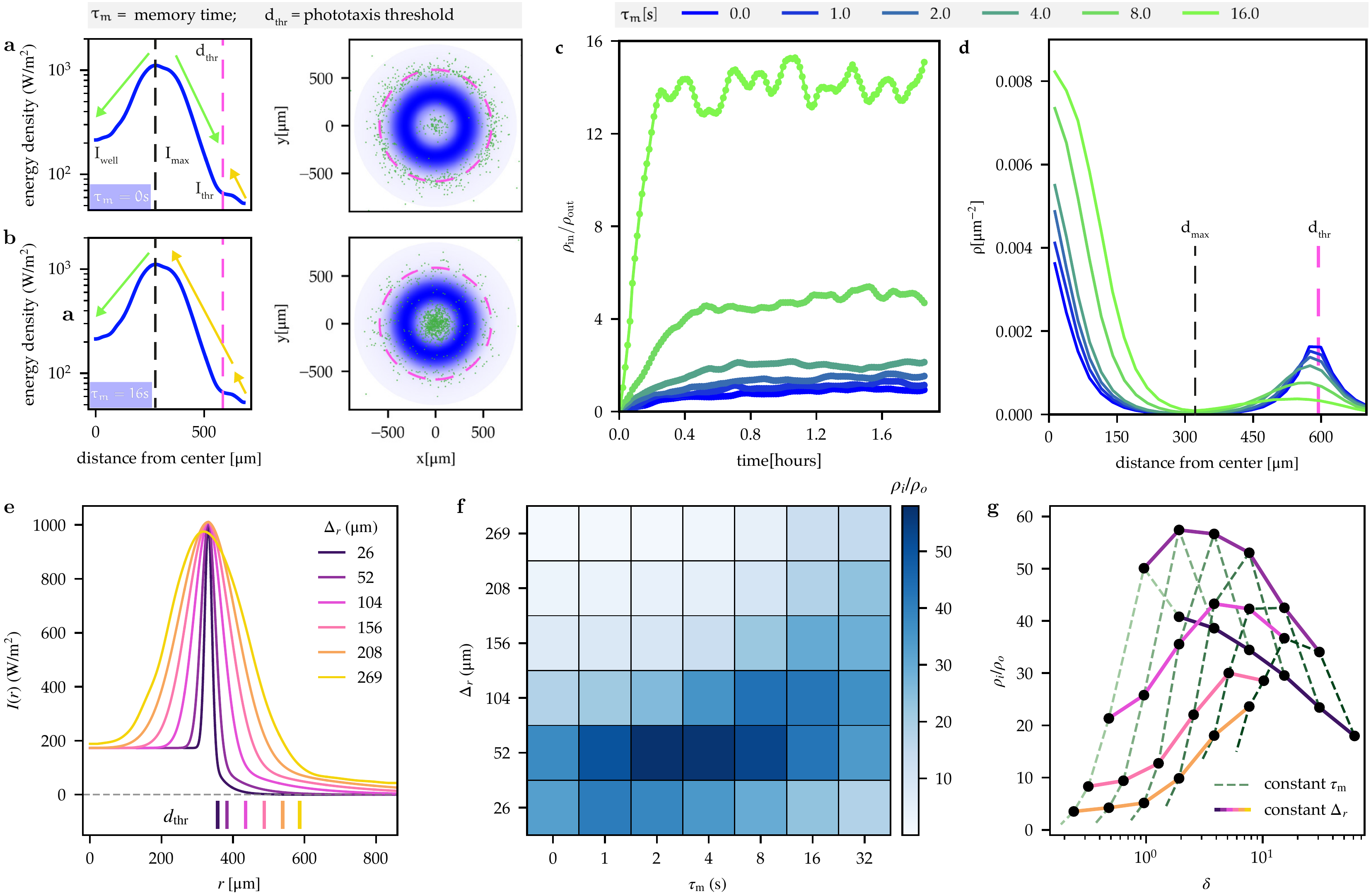}
    \caption{
    \textbf{Role of short-time memory in microalgal accumulation.} 
    \textbf{a–b)} Numerical modelling demonstrates that a short-time memory is needed for effective accumulation, illustrated through schematic (left) and simulation snapshots (right) panels.
    a) Without memory ($\tau_m = 0$), cells switch phototactic behaviour instantaneously upon crossing the intensity threshold ($d_{\text{thr}}$), leading to prolonged residence near the threshold rather than accumulation inside the ring.
    b) When memory is present ($\tau_m = \SI{16}{\second} $), cells integrate past light exposure, enabling them to cross the threshold and reach the location of maximum intensity (at $d_{\rm max}$) before switching to negative phototaxis and entering the ring.
    In the schematics, orange arrows indicate positive phototaxis, green arrows negative. 
    Simulation snapshots show the spatial distribution of the cells after one hour.
    \textbf{c)} The ratio of cell number density inside versus outside the ring, $\rho_{\text{in}}/\rho_{\text{out}}$, increases over time in a memory-dependent manner, with longer memory times leading to stronger accumulation. 
    \textbf{d)} The corresponding steady-state spatial profiles, measured after one hour of illumination, show that for short memory \rjj{durations}, cells predominantly accumulate near the intensity threshold $d = d_{\text{thr}}$; as memory increases, this peak decreases and the density within the ring becomes more pronounced.
    Calculations were run using a ring illumination with $r_{in} = \SI{150}{\micro\meter}$ and $r_{out} = \SI{420}{\micro\meter}$. \textbf{e–g)} Sweep in the parameter space $(\Delta_r, \tau_m)$. Panel \textbf{f)} shows a heatmap of the steady-state ratio $\rho_i/\rho_o$ as a function of the memory time $\tau_m$ and the distance $\Delta_r=d_{\rm thr}-d_{\rm max}$, obtained using the intensity profiles shown in panel \textbf{e)} (see Methods for the analytical definition of these profiles). \textbf{g)} Same data as in \textbf{f)} as a function of the dimensionless parameter $\delta=v_0\tau_m/\Delta_r$. Curves at constant $\Delta_r$ are shown with solid lines with the same colorcode as in \textbf{e)}, while curves at constant $\tau_m$ are shown with green dashed lines ($\tau_m$ increasing from light to dark green). Source data are provided as a Source Data file.
    }
    \label{fig4}
\end{figure*}

Although the switch in phototactic sign demonstrated in \autoref{fig3} is essential for driving accumulation, an instantaneous transition from positive to negative phototaxis at a local intensity threshold $I_{\rm thr}$ would not suffice to explain the observed localisation within the ring. In such a case, cells would accumulate near the isointensity contour corresponding to the switching threshold $I_{\rm thr}$, rather than inside the ring (\autoref{fig4}\textcolor{mygreen}{a}). 
This implies that any mechanism capable of producing the observed localisation must be non-instantaneous to allow the transition from positive to negative phototaxis to happen only after cells have reached the location of the intensity peak.

We therefore introduce a minimal theoretical model in which cells possess a short-time memory, allowing their behavioural response to depend not only on the instantaneous light intensity but also on its recent history, and which quantitatively reproduces the experimental observations. In the model, cells adjust their phototactic response based on an internal variable that integrates the light intensity over a finite memory time $\tau_{\rm m}$ (see Methods for details). When memory is absent, cells respond instantaneously to the local intensity and therefore accumulate predominantly near the threshold contour where the phototactic sign switches ($d_{\rm thr}$ in \autoref{fig4}\textcolor{mygreen}{a}). Introducing a short-time memory alters this behaviour: cells cross the switching region without reorienting immediately and continue towards the maximum intensity of the ring, where negative phototaxis subsequently redirects them toward the centre (\autoref{fig4}\textcolor{mygreen}{b}). Robust trapping within the intensity well then follows from the condition $I_{\rm well}>I_{\rm thr}$, which locks cells in the negatively phototactic state and prevents their escape.

We further quantified the role of memory by examining the time evolution of the numerical density ratio between the inside and outside of the ring in the model ($\rho_{\text{in}} / \rho_{\text{out}}$ in \autoref{fig4}\textcolor{mygreen}{c})\rj{, starting with no particle inside}. In the absence of memory, the ratio remains close to one over time, indicating no significant accumulation. When memory is included, the model predicts a progressive accumulation, reflecting the migration and retention of microalgae inside the ring. This behaviour agrees with the dynamics observed experimentally in \autoref{fig2}\textcolor{mygreen}{a}, \rj{suggesting} that a short-time memory accounts for the time-dependent dynamics of the phototactic response. The effect of memory is also evident in the steady-state radial density profiles ($\rho(r)$ in \autoref{fig4}\textcolor{mygreen}{d}). Without memory, cells accumulate near the radius $d_{\text{thr}}$ where the phototactic sign switches, with the central region \rj{only} sparsely populated. As the memory time increases, a peak emerges in the density profiles, reflecting accumulation in the central region of the ring, past the location of the intensity maximum. This trend is consistent with the experimental density profiles shown in \autoref{fig2}\textcolor{mygreen}{b}, highlighting memory as a key determinant of the spatial organisation. Additional simulations in a thinner ring (\textcolor{mygreen}{Figure S2}) show stronger accumulation at short memory times, as \rj{stochasticity in cell motion} more easily brings cells past the phototactic threshold inside the \rj{ring}.

The spatial distribution of cells is governed not only by the geometry of the illumination but also by its intensity, which determines whether and where the phototactic switching threshold is crossed within the light profile (\textcolor{mygreen}{Figure S3}\textcolor{mygreen}{a}).
In particular, when the threshold $I_{\rm thr}$ lies below the well intensity $I_{\rm well}$, 
cells enter the ring before switching to negative phototaxis, leading to accumulation in the centre, as shown in \autoref{fig2}, \autoref{fig4}, and \textcolor{mygreen}{Figure S3}\textcolor{mygreen}{a}-light green curve.
As light power is decreased, the threshold intensity $I_{\rm thr}$ becomes larger than $I_{\rm well}$ and closer to the peak intensity $I_{\rm max}$, and as a result the maximum cell number density inside the ring moves radially outward, until reaching $d_{\rm max}$ when $I_{\rm thr}>I_{\rm max}$. This shift is clearly observed in both simulations and experiments (\textcolor{mygreen}{Figure S3}\textcolor{mygreen}{b–c}), showing that the relative position of the switching point with respect to the intensity peak (i.e. $\Delta_r=d_{\rm thr}-d_{\rm max}$) shapes the accumulation pattern.

To further investigate how accumulation depends on the interplay between phototactic memory and the geometry of the illumination pattern, we have performed simulations by systematically varying the thickness of the ring (\autoref{fig4}\textcolor{mygreen}{e}; see Methods for the analytical definition of the intensity profiles) and the memory time. As shown in \autoref{fig4}\textcolor{mygreen}{f}, there actually exists an optimum memory time $\tau_m$ for a given ring thickness, which increases as $\Delta_r$ increases. Indeed, for a given ring thickness, memory must be large enough to ensure cells will reach the location of the maximum intensity before switching to negative phototaxis. However, if memory is too large, an individual agent remains for some time at the location of the maximum intensity before switching to negative phototaxis, limiting therefore the number of particles inside the ring. This is illustrated by the spatial distributions of the agents (at the steady-state) for $\tau_m=16$s and different ring thicknesses (\textcolor{mygreen}{Figure S4}\textcolor{mygreen}{a}): for small $\Delta_r$ there is a clear shoulder appearing at $d_{\rm max}$, because the agents spend some time diffusing around the location of maximum intensity before switching to negative phototaxis and going down the gradient within the ring. The characteristic time-scale for accumulation $\tau_a$ is itself monotonous with $\Delta_r$ and $\tau_m$: the wider the ring or the shorter the memory, the longer it takes to reach a steady-state (\textcolor{mygreen}{Figure S4}\textcolor{mygreen}{b}).

Since the level of accumulation depends non-trivially on the parameters, we now aim to bring some physical insights by introducing the dimensionless parameter $\delta=v_0\tau_m/\Delta_r$ which compares the memory length $v_0 \tau_m$ ($v_0$ being the self-propulsion speed) to the relevant distance to cross before switching phototactic sign $\Delta_r$. In the absence of noise and tumbling effects we expect accumulation for $\delta>1$, such that cells initiate the switch to negative phototaxis only after reaching the maximum intensity $I_{\rm max}$, where their movement down the gradient directs them towards the centre. As illustrated in \autoref{fig4}\textcolor{mygreen}{g}, we indeed observe that the accumulation ratio $\rho_i/\rho_o$ starts to increase at $\delta\sim 1$. However, for narrow rings (small $\Delta_r$) a significant accumulation plateau appears at low $\delta$ (i.e. low memory $\tau_m$). We attribute this to the stochasticity in the run-and-tumble motion, which makes it significantly probable to cross the short light barrier even when agents have already switched to negative phototaxis before reaching the intensity maximum, as a consequence of trajectory persistency.

Another effect of stochasticity is evident when varying the self-propulsion speed of the agents for a given ring thickness, \textcolor{mygreen}{Figure S5}\textcolor{mygreen}{a}. Interestingly here accumulation appears to be \rj{non-monotonic} with a maximum around \SI{30}{\micro\metre\per\second}, beyond which it decreases. This decrease arises from stochastic fluctuations in cell motion: faster and more persistent swimmers travel longer between reorientations, making them more likely to traverse the whole width of the ring in a few runs, even in the negative phototactic state, and escape the illuminated region (see trajectories for $v_0=50$ and \SI{90}{\micro\metre\per\second} in \textcolor{mygreen}{Figure S4}\textcolor{mygreen}{c-d}), thereby reducing the overall trapping efficiency.
This interpretation is further supported by the observation that curves corresponding to different \rj{speeds} tend to converge when the persistence length is matched by adjusting the tumbling rate (\textcolor{mygreen}{Figure S5}\textcolor{mygreen}{b}). 
These results clarify how the effectiveness of accumulation depends jointly on the light field geometry and the cell’s motility and memory timescales.

The key role of gradient reversal in enabling accumulation is further underscored by comparing the ring geometry to other illumination profiles. In particular, we experimentally tested two monotonic intensity landscapes: a one-dimensional line and a two-dimensional Gaussian profile (\textcolor{mygreen}{Figure S5}). In both cases, cells initially migrate up the gradient under positive phototaxis, but after switching to negative phototaxis, they are redirected down the same monotonic gradient—away from the illumination peak. As a result, no significant accumulation occurs near the centre. This contrast highlights that memory alone is not sufficient to produce the accumulation observed in the ring geometry: the geometry of the light field must also provide a spatial structure in which the phototactic gradient reverses direction, as in the ring.

In order to be as realistic as possible, steric interactions and collision-induced tumbling have been implemented in these simulations.   To assess the impact of cell–cell interactions on accumulation, we also compared simulations with and without
specific interaction terms. 
Even in the absence of interactions, accumulation persists and is stronger, confirming that phototactic memory alone
is sufficient to drive localisation (\textcolor{mygreen}{Figure S6}\textcolor{mygreen}{a}). Adding steric repulsion and collision-induced tumbling progressively reduces accumulation by limiting cell motion and promoting dispersal. In contrast, including motility quenching—where local crowding reduces swimming speed—enhances accumulation, consistent with the mechanism of Motility-Induced Phase Separation \cite{catesMotilityInducedPhaseSeparation2015}. 
However, this also leads to more ordered configurations inside the ring (\textcolor{mygreen}{Figure S6}\textcolor{mygreen}{b}), unlike the experiments where microalgae remain motile and disordered (Movie S1). Overall, these results demonstrate that accumulation is a consequence of single-cell behaviour, while its efficiency can be modulated via cell-cell interactions.

To complement the experiments and simulations, we developed an analytical description of phototactic microswimmers with memory. Cells are modelled as non-interacting active Brownian particles (ABPs) of speed $v_0$, undergoing translational and rotational diffusion \cite{bechingerActiveParticlesComplex2016}. As in the numerical approach, their orientation aligns or anti-aligns with the local light gradient, with the sign of the response determined by the average light intensity experienced over a finite memory time $\tau_{\rm m}$. This temporal integration renders the orientation dynamics non-local and prevents a closed equation for the cell number density $\rho(r)$ at arbitrary $\tau_{\rm m}$ (all details are provided in the Supplemental Information). Analytical tractability is achieved by performing a systematic expansion in the short-memory limit, assuming that $\tau_{\rm m}$ is small compared to the other timescales of the problem \cite{Kranz2016, Gelimson2016, Kranz2019}. At steady state, the swimmer distribution can be written as $\rho(r)\propto \exp[-U_{\rm eff}(\mathbf r)/k_BT]$, where the effective potential $U_{\rm eff}(\mathbf r)$ depends explicitly on the illumination profile, the phototactic threshold $I_{\rm thr}$, and the memory time $\tau_{\rm m}$. For vanishing memory ($\tau_{\rm m}=0$), particles accumulate at the threshold intensity $I_{\rm thr}$, as expected. With finite memory, however, the isointensity contour $I(r)=I_{\rm thr}$ loses stability once $\tau_{\rm m}$ exceeds a critical value $\tau_c$, which scales as $1/v_0$ and is consistent with the critical memory times extracted from simulations (\textcolor{mygreen}{Figure S9}). Even though the analytical model is derived in the short-memory regime, it highlights how memory fundamentally alters the coupling between active particles and external fields: introducing memory, even at short times, makes this coupling nonlinear and history-dependent, leading to qualitatively new accumulation regimes such as the central localisation observed here.

\subsection*{Long-term phototactic adaptation}
The experimental accumulation dynamics should not be interpreted as convergence to a steady state but rather in terms of distinct time regimes:
accumulation is detectable within minutes and reaches a broad maximum after approximately one hour (\autoref{fig2}\textcolor{mygreen}{a}), after which the density inside the ring gradually declines over several hours (\textcolor{mygreen}{Figure S7}\textcolor{mygreen}{a}).
While the inclusion of memory in the model successfully captures the short-time dynamics of accumulation, it always predicts a steady accumulation plateau (\autoref{fig4}\textcolor{mygreen}{c}).
This discrepancy cannot be accounted for by tuning the model’s parameters, as all fixed-parameter configurations ultimately lead to stationary density profiles at long times. \rj{Instead, the experimentally observed decline is seemingly consistent with an adaptation dynamic leading to a gradual reduction in phototactic sensitivity from sustained light exposure \cite{Arrieta2017}, operating on a timescale of tens of minutes}. Modelling this effect as a temporal decay in the capacity of cells to align (or anti-align) with the gradient (see Methods) reproduces the observed decrease in accumulation (\textcolor{mygreen}{Figure S7}\textcolor{mygreen}{b}). Together, these results indicate that memory governs the initial accumulation, while adaptation controls the long-time spatial organisation.

\section*{Discussion}
In summary, our study demonstrates that structured light landscapes can guide microalgae populations and localise them in defined spatial regions. By finely analysing single-cell dynamics within ring-shaped illumination patterns, we have uncovered two key features of their phototactic behaviour. Firstly, while the response of \textit{C.~reinhardtii} to light intensity gradients has already been established \cite{Brunet2022, Arrieta2017}, we have shown here that cells actively steer along intensity gradients by increasing their run length when aligned (or anti-aligned) with the stimulus. \rjj{While the exact steering mechanism is probably altered by the strong confinement of our setup, trapping still takes place in unconfined situations (Movie S2), which calls for investigation of the 3D motion of the algae along light gradients.}  Secondly, a comparison with a minimal theoretical model shows that a short-term memory of light exposure is necessary to reproduce the observed accumulation. The gradual decrease in cell number density over time further suggests the onset of long-term adaptation, pointing to a layered phototactic response in which short-time memory governs navigation while adaptation modulates behaviour under prolonged exposure \cite{Arrieta2017}.
More broadly, our results illustrate how history-dependent sensing can shape active dynamics in structured environments. Temporal sensing can be converted into a spatial displacement, so that population organisation is set jointly by field geometry, motility persistence and the internal response time. Our experiments provide a realisation of this principle, which may extend to other taxis-driven microorganisms, and inspire the design of synthetic microswimmers and active materials endowed with some basic intelligence (memory, programmability, adaptability) and controlled by external fields \cite{ju_technology_2025}, for which control strategies have remained relatively limited so far \cite{velluvakandy_artificial_2026,dey_oscillatory_2022}.

Our observations open up several avenues for further investigation into the phototactic response of \textit{C.~reinhardtii}. While run-length modulation in response to external gradients is a well-established navigation strategy in chemotactic bacteria, the mechanisms that enable \textit{C.~reinhardtii} to respond to light intensity gradients remain largely unexplored. In particular, identifying the sensory and molecular pathways involved in this behaviour is a compelling direction for future work. It is unknown whether this gradient-based response involves the eyespot—typically associated with directional alignment to light—or whether it reflects an alternative or complementary sensory mechanism. Secondly, our memory model—based on the temporal integration of past light intensities—does not reflect the biological implementation of phototactic memory. Cells are instead likely to rely on the dynamics of internal variables that couple to light, as proposed by Laroussi \textit{et al.}~\cite{Laroussi2024}. Implementing memory using this intracellular-species model (see Supplemental Information, “S model”) also results in accumulation within the ring and reproduces both the steady-state accumulation and the gradual long-term decrease in density (\textcolor{mygreen}{Figure S7}\textcolor{mygreen}{c}).
Finally, imposing a simple delay on the response also leads to accumulation which is quantitatively the same as the memory model (see \textcolor{mygreen}{Figure S7}\textcolor{mygreen}{d}). Extending our approach to other illumination geometries and dynamic light conditions could provide insights into how intensity gradients are sensed, how light signals are processed internally, and how the switch in phototactic sign is controlled. Such experiments would complement the existing knowledge on the molecular details of the phototactic response \cite{hegemann_algal_2008,kreimer_green_2009,bohm_channelrhodopsin-1_2019,baidukova_gating_2022,wolfram_multifactorial_2023,wolfram_insights_2024} and help to establish a comprehensive picture of phototactic navigation. Beyond biological contexts, the mechanisms uncovered here could inform the design of light-controlled micro-swimmer systems and offer routes to programmable transport and localisation in synthetic active matter.

\section*{Methods}
\subsection*{\label{sec:ExperimentsPrep}Culture preparation}
Wild-type CC125 strain was obtained from the Chlamydomonas Resource Center.
Cells were cultured in Tris-Acetate-Phosphate (TAP) medium \cite{alma990094274380107871} under an illumination of \SI{70}{\micro \mole \per\square\metre\per\second} at \SI{22}{\celsius}, on an orbital shaker operating at 160 rpm. 
To synchronise the population, cells were exposed to a day/night light cycle of 16/8 hours.
Cells were harvested during the exponential growth phase (concentration between approximately \SI{0.4 \times~10^{6}}{\per\milli\litre} and \SI{5 \times~10^{6}}{\per\milli\litre}) and injected into a microfluidic chamber with a thickness of approximately \SI{20}{\micro\metre}, thereby confining the suspension to a quasi-two-dimensional geometry. To minimise the variability associated with circadian regulation of phototaxis, cells were harvested and experiments were performed in the morning, at the same stage of the light cycle for all experiments.

\subsection*{\label{sec:ExperimentsSetup}Experimental setup}
\textcolor{mygreen}{Figure S1} shows a schematic of the experimental setup. A laser (Cobolt, 06-MLD, \SI{473}{\nano \meter}, \SI{330}{\milli \watt} of power) and an axicon (Thorlabs, AX255-A) are used to illuminate the microalgae with a ring-shaped light profile. 
The laser is directed to the sample via two mirrors (M1, M2) illuminating the back aperture of a 10x objective (Nikon, N10X-PF).
By changing the position of the sample and the illumination objective it is possible to control both the inner and outer diameter of the ring illumination. 
In particular, in \autoref{fig2} the outer diameter was increased by using a 4x (Olympus, PLN 4X) instead of a 10x objective. The illumination profiles were measured at the sample plane, where the microalgae are confined and imaged. The images shown in Fig. \autoref{fig2}a corresponds therefore to the light intensity the cells sample along their trajectories through their cellular photosensors.
To create a Gaussian illumination, we removed the axicon to image the beam on the sample. The illumination envelope was fixed to be close to the size of the ring illumination. 
The intensity of the laser—controlled by a half-waveplate (FOCtek, WPF212H) and a polarising beam splitter (FOCtek, BSC1204) was then adjusted to have a comparable energy density in the different optical landscapes. 
The microalgae were imaged with a 20x objective (Nikon, N20X-PF) and a tube lens (Thorlabs, LA1805-A-ML) on a camera (Basler, acA5472-17um).  A red LED (Thorlabs, \SI{660}{\nano \meter}, M660F1), for simplicity not depicted in \textcolor{mygreen}{Figure S1}, coupled to a fibre (Thorlabs, M28L01) is used for imaging. This wavelength was chosen to avoid any interference with the phototactic response. The algae dynamics was analysed by reconstructing their trajectories from videos (typically \rjj{of 2 minutes duration at 10 fps}) using a homemade code based on the python package \textit{Trackpy} \cite{crockerMethodsDigitalVideo1996,allan_daniel_b_2023_7670439}.
The position of the sample was controlled in three dimensions using an \textit{xyz}-stage (Thorlabs, RB13M/M). The \textit{z}-coordinate was adjusted with a stepper motor (Thorlabs, ZFS25B).

\subsection*{\label{sec:Numerics}Numerical model}
 \label{Model1}
We base our model for the algae on the active Brownian particle model ({\it ABP}) in two dimensions \cite{jacucciPatchyEnergyLandscapes2024}:
\begin{align}
    \dot{\textbf{r}}_i(t) &=v_0\textbf{u}_i(\phi_i)-\frac{1}{\gamma}\sum_{j\neq i}\boldsymbol\nabla V\left(\left|\textbf{r}_i-\textbf{r}_j\right|\right)+\sqrt{2D}\boldsymbol{\eta}_i(t), \label{eq:position}\\
   \dot{\phi}_i(t) &= \omega(\textbf{r}_i,\phi_i,\mu_i)+\sqrt{2D_r}\eta_i^r (t), \label{eq:angle}
\end{align}
where $\textbf{r}_i$ is the particle's position, $\phi_i$ its orientation angle, $\boldsymbol{\eta}_i$ and $\eta_i^r$ are respectively translational and rotational white noise terms, $v_0$ is the active propulsion speed, $\textbf{u}_i=(\cos(\phi_i),\sin(\phi_i))$ its direction, $\gamma$ is the friction coefficient, $D$ and $D_r$ are respectively the translational and rotational diffusion coefficients and $V$ is a repulsive inter-particle Weeks-Chandlers-Andersen potential. In order to introduce phototaxis/antiphototaxis and memory, we integrate this model with run-and-tumble events, represented by the term $\omega$. These events are distributed in time with a Poisson process: at each time step $\Delta t$ the probability of tumbling is $P_t=1-\text{e}^{-\Delta tR_t}$, where $R_t$ is the tumbling rate. The term $\omega$ is then defined as
\begin{equation}
    \omega(\textbf{r}_i,\phi_i,\mu_i)=\begin{cases}
        \zeta_a\delta(t)\qquad \text{for }\zeta_r<P_t\\
        0\qquad\qquad \text{for }\zeta_r\geq P_t
    \end{cases},
\end{equation}
where $\zeta_a, \zeta_r$ are random variables uniformly distributed on $[0,2\pi]$ and $[0,1]$ respectively. Now, the fundamental part of this model is how $R_t$ is determined, as it is dependent on multiple factors: the positions of the particles $\textbf{r}_i$, their orientations $\phi_i$ and their internal state $\mu_i$. We encode the effects of the light intensity field $I(\mathbf r)$ through the
tumbling rate
\begin{equation}
\lambda(\mathbf r_i,\phi_i,\mu_i)=
\begin{cases}
C,\\
\lambda_0(\mathbf r_i)\!\left[
1+\mu_i(t)\,
g_R\!\big(\boldsymbol\nabla I(\mathbf r_i)\!\cdot\!\mathbf u_i(\phi_i)\big)
\right].
\end{cases}
\label{eq:Rt}
\end{equation}

\noindent
The upper branch applies when a collision occurs, defined by the existence of
a particle $j\neq i$ such that $\lvert\mathbf r_i-\mathbf r_j\rvert<\sigma$;
otherwise the tumbling rate is controlled by the local light field.
The baseline tumbling rate is set by the intensity-dependent function
$\lambda_0(\mathbf r_i)=f_R(I(\mathbf r_i))$, where $f_R$ is a positive,
monotonically increasing function extracted from experiments
(\textcolor{mygreen}{Figure S8}).
The function $g_R$ quantifies how the alignment between the propulsion
direction $\mathbf u_i$ and the local intensity gradient
$\boldsymbol\nabla I$ biases tumbling.
Its functional form is determined by the experimental data in \autoref{fig3}: fitting the average tumbling rate $\lambda(\theta)$ (the inverse of the average run time shown in \autoref{fig3}a) by an affine function $\lambda(\theta)=\lambda_0(1+\alpha\theta)$ leads to $g_R(\cos{(\theta)})=\alpha\theta$, and therefore $g_R(x)=\alpha \arccos{(x)}$. The sign of the response (positive or negative phototaxis) is itself controlled by the value (i.e. $\pm 1$) of the internal state $\mu_i$, which is defined as
\rjj{
\begin{equation}
    \mu_i(t)=\text{sign}\left[\int_{t-\tau_\text{m}}^t\text{d}t' (I_\text{thr}-I(\textbf{r}_i(t')))\right],
    \label{mu}
\end{equation}
}
\rjj{where $I_{\rm thr}$} is the intensity threshold above which algae become antiphototactic and $\tau_\text{m}$ is the memory time. 
The value of \(I_{\rm thr}\) was not fixed directly from the experiments. The simulated density profiles exhibit a secondary peak at the threshold contour (\autoref{fig4}\textcolor{mygreen}{d} and \textcolor{mygreen}{Figure S2}), which is not resolved experimentally (\autoref{fig2}\textcolor{mygreen}{b}). This peak is probably smoothed out by cell-to-cell variability in the phototactic response (e.g. distribution of $I_{\rm thr}$ among the population) and slight asymmetries in the illumination profile (deviation from axisymmetry notably). The precise value of \(I_{\rm thr}\) is therefore not critical for the localisation mechanism: accumulation inside the ring is observed whenever \(I_{\rm thr}\) lies below the maximum intensity of the light field $I_{\rm max}$, as shown in \textcolor{mygreen}{Figure S3}, while strong trapping and central localisation occurs when $I_{\rm thr}<I_{\rm well}$.
$\mu_i$ effectively switches the effect of $g_R$ from phototactic to antiphototactic after the algae has spent enough time in a region where $I>I_\text{thr}$ and vice versa. Finally, we have a collision term $C$ that sets $R_t=C$, arbitrarily large, if a particle $j\neq i$ exists such that $|\textbf{r}_i-\textbf{r}_j|<\sigma$.

 \textcolor{black}{As a variant of our model for memory, we also considered a simple delay effect for the switching of phototaxis. This was done by defining $\mu_i(t)$ as follows:
 \rjj{
\begin{equation}
    \mu_i(t)=\text{sign}[I_\text{thr}-I(\textbf{r}_i(t-\tau_{\rm d}))].
    \label{mu}
\end{equation} }
}
In order to reproduce the long-time decay in accumulation observed in experiments, we introduce a time dependence for \textcolor{black}{the gradient alignment $g_R$}. This models the ageing of the system by \rjj{exponential decay} over time with the following relation:
\begin{equation}
    \textcolor{black}{ g_R(\boldsymbol{\nabla} I(\textbf{r}_i)\cdot\textbf{u}_i(\phi_i),t)=g_R(\boldsymbol{\nabla} I(\textbf{r}_i)\cdot\textbf{u}_i(\phi_i))}\text{e}^{-tr_g},
\end{equation}
where $r_g$ defines the decay rate.

For the sweep in the parameter space $(\Delta_r, \tau_m)$, we used an analytical expression for the intensity profiles (\autoref{fig4}\textcolor{mygreen}{e}) in order to systematically vary  $\Delta_r$ in a controlled way. The intensity profiles were defined through: 
\begin{equation}
I(r)=
\begin{cases}
A\exp{\Big(-\frac{(r-d_{\rm max})^2}{2s^2}\Big)}+a,~r\leqslant d_{\rm max}\\
A\exp{\Big(-\frac{(r-d_{\rm max})^2}{2s^2}\Big)}+a\exp{\Big(\frac{d_{\rm max}-r}{b}}\Big),~r>d_{\rm max}
\end{cases}
\label{intensity_profiles}
\end{equation}
with $A=(I_{\rm max}-I_{\rm well})/(1-\exp{(-d_{\rm max}^2/2s^2)})$, $a=I_{\rm max}-A$ and $b=348 {\rm \mu m}$. The distance $\Delta_r$ was then varied by tuning the width of the Gaussian $s$.

\section*{Data availability}
Source data are provided with this paper. Further data supporting the findings of this article are openly available \cite{FigShare}.

\section*{Code availability}
The code used to perform the numerical simulations is publicly available \cite{FigShare}. 



\putbib

\section*{Funding}
This project has received funding from the Alliance Sorbonne Universit\'e under the Emergence 2023 scheme (No ANR-IDEX-00040$\_$2, X23JR31001) and from the ANR JCJC 2023 (No ANR-23-CE30-0009-01).

\section*{Author contributions}
Author contributions are defined based on the CRediT(Contributor Roles Taxonomy) and listed alphabetically. Conceptualization: J-FA, SG, GJ, RJ. Data curation: DB, GJ. Formal analysis: DB, PI, GJ. Resources: J-FA, SG, GJ, RJ, LT. Writing: GJ and RJ wrote the manuscript with the help of all authors. Funding acquisition: J-FA, SG, GJ, RJ

\section*{Competing interests}
The authors declare no competing interests.

\end{bibunit}

\newpage
\textcolor{white}{blabla}
\newpage

\setcounter{figure}{0}
\renewcommand{\thefigure}{S\arabic{figure}}
\setcounter{equation}{0}
\renewcommand{\theequation}{S\arabic{equation}}%

\newcommand{\nocontentsline}[3]{}
\newcommand{\tocless}[2]{\bgroup\let\addcontentsline=\nocontentsline#1{#2}\egroup}

\onecolumn

\begin{bibunit}[naturemag-doi]
\begin{center}
  {\fontsize{22pt}{26pt}\selectfont\textbf{Supplementary Information}}\\[0.75em]
  {\fontsize{16pt}{20pt}\selectfont Controlling microalgae populations by phototactic memory}
\end{center}
\vspace{3em}

\section*{S model}
Taking inspiration from the model for \textit{C.~reinhardtii} described in Ref. \cite{Laroussi2024}, we modified the model used in the main text, and more specifically the functioning of the internal state variable $\mu_i$ (Eq.~\ref{mu}). The new internal state variable, $\mu_i^s$, depends on a new variable $s_i$ and reads as follows:
\begin{equation}
    \mu_i^s=\text{sign}(s_T-s_i),
\end{equation}
where $s_i\in[0,1]$ represents the activation percentage within the algae of a biochemical species able to reverse phototaxis after enough of it is activated; a threshold given by the parameter $s_T$. The dynamics of $s_i$ is determined by the differential equation
\begin{equation}
    \dot{s}_i=\gamma^sI(\textbf{r}_i)(1-s_i)-s_i/\tau^s,
\end{equation}
where $\gamma^s I(\textbf{r}_i)$ is the rate at which inactive species become active, $I(\textbf{r}_i)$ is the light intensity in position $\textbf{r}_i$ and $\tau^s$ is the typical time at which the species decays back into its deactivated stage. This is another effective way of introducing memory into our system.

\section*{Analytical model of microalgae phototaxis with memory}
We consider an active Brownian particle (ABP), whose position and orientation at time $t$ are denoted by $\rr(t)$ and $\nn(t)$, respectively. The calculation will be valid in dimension 2 and 3, and we denote by $d$ the spatial dimension. The propulsion speed of the ABP is denoted by $v_0$, and its `bare' or passive diffusion coefficient is denoted by $D$. The position $\rr$ therefore obeys the equation:
\begin{equation}
	\frac{\dd \rr}{\dd t} = v_0 \nn(t) + \sqrt{2D} \boldsymbol{\xi}(t),
\end{equation}
where $\boldsymbol{\xi}(t)$ is a Gaussian white noise of zero average: $\langle \xi_i(t) \rangle =0$, and unit variance:  $\langle \xi_i(t)\xi_j(t') \rangle =\delta_{ij} \delta(t-t')$.

We assume that the orientation of the ABP fluctuates, with a orientational diffusion coefficient $D_r$ (homogeneous to an inverse time), and responds to the underlying light pattern $I(\rr)$, by either aligning or  anti-aligning with the gradient $I(\rr)$. In order to construct the equation satisfied by $\nn$, we first recall that if a unit vector $\uu$  obeys the differential equation: $\frac{\dd \uu}{\dd t} = (\uu\times\boldsymbol{a})\times \uu $, then it will align with $\boldsymbol{a}$ (it may also anti-align with $\boldsymbol{a}$ but this is an unstable state).

Consequently, the evolution equation for $\nn$ must include a term proportional to $[\nn\times \nabla I(\rr(t))] \times \nn$. The sign of this term must be positive (resp. negative) if the swimmer is in a state of positive (resp. negative) phototaxis. Therefore, we introduce a function threshold function $f(x)$ that is typically negative for $x<\mathcal{I}_\text{thr}$, and positive otherwise (see a sketchy representation on Fig. \ref{threshold_function}). This function is evaluated at the average intensity profile seen on the time interval $[t-\tau_\text{m},t]$, where $\tau_\text{m}$ is the memory time:
\begin{equation}
	\mathcal{I}(t) = \int_{-t-\tau_\text{m}}^t \dd t' \; I(\rr(t')).
\end{equation}

The position and orientation of the ABP therefore obey the coupled overdamped Langevin equations:
\begin{eqnarray}
	\frac{\dd \rr}{\dd t} &=& v_0 \nn(t) + \sqrt{2D}\boldsymbol{\xi}(t) \label{Langevin_r}\\
	\frac{\dd \nn}{\dd t} &=& f(\mathcal{I}(t)) [\nn(t)\times \nabla I(\rr(t))] \times \nn(t) + \sqrt{2D_r}\boldsymbol{\zeta}(t)  \label{dndt}\\
	&=& f \left(  \frac{1}{\tau} \int_{t-\tau_{\rm m}}^t \dd t'\; I(\rr(t')) \right) [\nn(t)\times \nabla I(\rr(t))] \times \nn(t) + \sqrt{2D_r}\boldsymbol{\zeta}(t) \label{Langevin_n}
\end{eqnarray}
where $f$ is a threshold function of the shape shown on Fig. \ref{threshold_function} (its expression does not need to be specified for now), and where $\boldsymbol{\zeta}(t)$ is a Gaussian white noise of zero average: $\langle \zeta_i(t) \rangle =0$, and unit variance:  $\langle \zeta_i(t)\zeta_j(t') \rangle =\delta_{ij} \delta(t-t')$. 

\subsection*{Fokker-Planck equation in the limit of short memory}

The equation obeyed by $\nn(t)$ is non-local in time, which makes the analysis of the Langevin Eqs. \eqref{Langevin_r} and \eqref{Langevin_n} complicated. To make progress, we follow the idea from Refs. \cite{Kranz2016, Gelimson2016, Kranz2019} and study the behavior of the ABP in the limit where it has a short memory, i.e. when the timescale $\tau_{\rm m}$ is typically smaller than all the other timescales of the problem. The average intensity $\mathcal{I}(t)$ is first rewritten using the change of variable $t'' \leftarrow t-t'$:
\begin{equation}
	\mathcal{I}(t) = \frac{1}{\tau} \int_{0}^{\tau_{\rm m}} \dd t''\; I(\rr(t-t'')).
	\label{average_I}
\end{equation}
Since $0\leq t'' \leq \tau_{\rm m}$ and since $\tau_{\rm m}$ is assumed to be small, the position of the ABP can be expanded as: 
\begin{equation}
	\rr(t-t'') \simeq \rr(t) - t'' v_0 \nn(t).
\end{equation}
Reinjecting in Eq. \eqref{average_I} yields, in the limit $\tau_{\rm m} \to 0$:
\begin{equation}
	\mathcal{I}(t) = I(\rr(t)) - \frac{\tau_{\rm m} v_0}{2} \nn(t) \cdot \nabla I(\rr(t)) + \mathcal{O}(\tau^2).
\end{equation}
Consequently, at linear order in $\tau_{\rm m}$, the threshold function reads:
\begin{equation}
	f(\mathcal{I}(t)) = f(I(\rr(t))) -  \frac{\tau_{\rm m} v_0}{2}  [\nn(t) \cdot \nabla I(\rr(t))]  f'(I(\rr(t))) +  \mathcal{O}(\tau^2)
\end{equation}
Within this approximation, the Langevin equations for $\rr$ and $\nn$ are then:
\begin{eqnarray}
	\frac{\dd \rr}{\dd t} &=& v_0 \nn(t) + \sqrt{2D}\boldsymbol{\xi}(t) \\
	\frac{\dd \nn}{\dd t} &=& \Big\{  f(I(\rr(t))) -  \frac{\tau_{\rm m} v_0}{2}  [\nn(t) \cdot \nabla I(\rr(t))]  f'(I(\rr(t)))  \Big\}[\nn(t)\times \nabla I(\rr(t))] \times \nn(t) + \sqrt{2D_r}\boldsymbol{\zeta}(t) 
\end{eqnarray}
These equations are now local in time, and are easier to integrate.

Introducing the angular velocity:
\begin{equation}
	\boldsymbol{\omega}[\rr(t),\nn(t)] \equiv  \Big\{  f(I(\rr(t))) -  \frac{\tau_{\rm m} v_0}{2}  [\nn(t) \cdot \nabla I(\rr(t))]  f'(I(\rr(t)))  \Big\}[\nn(t)\times \nabla I(\rr(t))],
	\label{angular_velo_omega}
\end{equation}
we get the following Fokker-Planck equation for $P(\rr,\nn,t)$, i.e. the probability to find the ABP at position $\rr$ and with orientation $\nn$ at time $t$:
\begin{equation}
	\partial_t P(\rr,\nn,t) = D \nabla^2 P(\rr,\nn,t)+ D_r \rotop^2 P(\rr,\nn,t) - v_0 \nabla \cdot (\nn P(\rr,\nn,t)) - \rotop\cdot [\boldsymbol{\omega}(\rr,\nn) P(\rr,\nn,t)]
	\label{FPeq}
\end{equation}
where $\rotop = \nn \times \partial_{\nn} $ is the orientational gradient operator \cite{Doi1988,Dhont1996}. The strategy is now to find the distribution of the position of the swimmer in the stationary limit, defined as:
\begin{equation}
	\rho(\rr) =	\lim_{t\to\infty}  \int \dd \nn \; P(\rr,\nn,t),
\end{equation}
and to determine how this stationary distribution of the microswimmer is related to the illumination pattern $I(\rr)$, to the threshold function $f$, and to the typical memory time $\tau_{\rm m}$.

\subsection*{Moment expansion}

We integrate Eq. \eqref{FPeq} over all orientations $\nn$, and find that it gives an unclosed equation for the quantity $\rho(\rr,t) \equiv \int \dd \nn \; P(\rr,\nn,t)$:
\begin{equation}
	\partial_t \rho(\rr,t) = D \nabla^2 \rho(\rr,t) - v_0 \nabla \cdot \TT (\rr,t) 
	= -\nabla \cdot [-D\nabla\rho(\rr,t) + v_0\TT (\rr,t) ]
	\label{eq_rho}
\end{equation}
where $\TT$ is the polarization field, defined as:
\begin{equation}
	\TT(\rr,t) \equiv \int \dd \nn \; \nn P(\rr,\nn,t).
\end{equation}
The next step is to obtain the evolution equation for $\TT$, which is expected to be unclosed, and to involve higher-order moments (nematic terms and so on): this is the usual `moment expansion' that arises in active matter theories \cite{Ahmadi2006,Golestanian2012,Brotto2013,Saha2014,Golestanian2022,Kurzthaler2016a, gautry2025closuresmomentexpansionanisotropic}.

Multiplying Eq. \eqref{FPeq} with component $n_i$ and integrating over $\nn$ yields the following equation for the $i$-th component of the polarization $\TT$ (repeated indices are implicitly summed over):
\begin{equation}
	\partial_t T_i(\rr,t) = D \nabla^2 T_i -2D_r T_i - v_0 \partial_j \int \dd \nn \; n_i n_j P(\rr,\nn,t) - \int \dd \nn \; n_i \mathcal{R}_j( \omega_j (\rr,\nn) P(\rr,\nn,t))
\end{equation}
The calculation of the last term is more complicated. Integrating by parts and using the general relation $\mathcal{R}_jn_i = -\varepsilon_{jik} n_k$ ($\varepsilon_{jik}$ being the Levi-Civita tensor), one gets
\begin{eqnarray}
	- \int \dd \nn \; n_i \mathcal{R}_j( \omega_j (\rr,\nn) P(\rr,\nn,t))
	&=&
	\varepsilon_{jik} \int \dd \nn \; n_k  \omega_j (\rr,\nn) P(\rr,\nn,t) \\
	&=& 	 \varepsilon_{jik} 	 \varepsilon_{jbc}  \int \dd \nn \; n_k P(\rr,\nn,t) 
	\left[f(I(\rr)) - \frac{\tau_{\rm m} v_0}{2} (n_a \partial_a I) f'(I)  \right] n_b \partial_c I
\end{eqnarray}
where we used the definition of $\boldsymbol{\omega}$ [Eq. \eqref{angular_velo_omega}] to get the second equality. It can be seen that this term yields higher-order moments, namely tensors of rank two and three:
\begin{eqnarray}
	\mathbf{Q}^{(2)}(\rr,t) &=& \int \dd \nn \;  \nn\nn P(\rr,\nn,t)\\
	\mathbf{Q}^{(3)}(\rr,t) &=& \int \dd \nn \;  \nn\nn\nn P(\rr,\nn,t)
\end{eqnarray}
They are approximated as follows, in order to close the hierarchy of equations \cite{Golestanian2012,Saha2014,Golestanian2022,Ahmadi2006}:
\begin{eqnarray}
	{Q}_{ij}^{(2)}(\rr,t) &\simeq& \frac{1}{d} \delta_{ij}  \rho(\rr,t),\\
	{Q}_{ijk}^{(3)}(\rr,t) &\simeq&  \frac{1}{d+2}( \delta_{ij}  T_k(\rr,t) +  \delta_{ik}  T_j(\rr,t)+ \delta_{jk}  T_i(\rr,t)).
\end{eqnarray}
This yields the following equation for $\TT$:
\begin{eqnarray}
	\partial_t \TT(\rr,t) &=& D \nabla^2 \TT -(d-1)D_r \TT - \frac{v_0}{d} \nabla \rho  +\frac{d-1}{d} f(I(\rr)) \rho \nabla I   +\frac{v_0\tau_{\rm m}}{2(d+2)} f'(I(\rr)) [(\nabla I)^2 \TT - d (\TT\cdot \nabla I) \nabla I].
\end{eqnarray}

We split the polarization field over a part which is parallel to the intensity gradient, and a part which is perpendicular. Denoting by $\uu = \nabla  I /|\nabla I|$ the local orientation of the gradient, and defining:
\begin{eqnarray}
	\TT_\parallel(\rr,t) &=& (\TT(\rr,t)\cdot \uu)\uu, \\
	\TT_\perp(\rr,t) &=& \TT(\rr,t)- (\TT(\rr,t)\cdot \uu)\uu, 
\end{eqnarray}
we get the following equations for $\TT_\parallel$ and $\TT_\perp$:
\begin{eqnarray}
	\partial_t \TT_\parallel(\rr,t) &=& D \nabla^2 \TT_\parallel -(d-1)D_r \TT_\parallel - \frac{v_0}{d} (\uu\cdot \nabla \rho)\uu  +\frac{d-1}{d} f(I(\rr)) \rho \nabla I - v_0 \tau_{\rm m}\frac{d-1}{2(d+2)} f'(I(\rr)) |\nabla I|^2 \TT_\parallel \\
	\partial_t \TT_\perp(\rr,t) &=& D \nabla^2 \TT_\perp - (d-1)D_r \TT_\perp - \frac{v_0}{d} (\nabla\rho-(\uu\cdot \nabla \rho)\uu)+ \frac{v_0 \tau_{\rm m}}{2(d+2)} f'(I(\rr)) |\nabla I|^2 \TT_\perp
\end{eqnarray}
In the stationary limit, and neglecting large scale variations of the polarisation field (i.e. $\nabla^2 \TT_\parallel \simeq 0$ and $\nabla^2 \TT_\perp \simeq 0$), we get:
\begin{eqnarray}
	\TT_\parallel(\rr,t) &\simeq & \frac{1}{(d-1)D_r + v_0 \tau_{\rm m}\frac{d-1}{2(d+2)} f'(I(\rr)) |\nabla I|^2} \left[ - \frac{v_0}{d} (\uu\cdot \nabla \rho)\uu  +\frac{d-1}{d} f(I(\rr)) \rho \nabla I\right] \label{Tpar_stat}\\
	\TT_\perp(\rr,t) &\simeq & \frac{1}{(d-1)D_r - \frac{v_0 \tau_{\rm m}}{2(d+2)} f'(I(\rr)) |\nabla I|^2} \left[ - \frac{v_0}{d} (\nabla \rho- (\uu\cdot \nabla \rho)\uu)\right] 	
\end{eqnarray}

In the stationary state, Eq. \eqref{eq_rho} gives: $-D\nabla \rho +v_0 \TT \simeq 0$. Assuming that the variations of polarizations are essentially controlled by the gradient of the light pattern, we write $\TT\simeq \TT_\parallel$, and using Eq. \eqref{Tpar_stat}, one gets:
\begin{equation}
	0 \simeq -D \nabla \rho +  \frac{v_0}{(d-1)D_r + v_0 \tau_{\rm m}\frac{d-1}{2(d+2)} f'(I(\rr)) |\nabla I|^2} \left[ - \frac{v_0}{d}  \nabla \rho\uu  +\frac{d-1}{d} f(I(\rr)) \rho \nabla I\right] 
\end{equation}
At linear order in $\tau_\text{m}$, we get:
\begin{equation}
	\frac{\nabla \rho}{\rho} = \frac{v_0 f(I(\rr))}{dD_r D_\text{act}} \left[1+\frac{v_0 \tau_{\rm m} D }{2(d+2) D_\text{act} D_r} f'(I(\rr)) |\nabla I|^2\right] \nabla I,
\end{equation}
where $D_\text{act}$ is the bare long-time diffusion coefficient of the ABP: $D_\text{act}=D + \frac{v^2}{d(d-1) D_r}$.

Finally, we refomulate the stationary state in terms of an effective potential, which is such that $\rho \sim \exp(-U_\text{eff}/\kB T )$, i.e. 
\begin{equation}
	\frac{\nabla \rho}{\rho} \sim -\frac{1}{\kB T}  \nabla U_\text{eff},
\end{equation}
where:
\begin{equation}
	U_\text{eff}(\rr) = - \kB T \int_{\rr} \dd \rr'\frac{v_0 f(I(\rr'))}{dD_r D_\text{act}} \left[1+\frac{v_0 \tau_{\rm m} D }{2(d+2) D_\text{act} D_r} f'(I(\rr')) |\nabla I(\rr')|^2\right] \nabla I(\rr')
\end{equation}

As a check, in the case where there is no memory and where the phototactic effect is `constant' (i.e. $\tau_{\rm m}=0$ and $f$ independent from $I$), we get:
\begin{equation}
	U_\text{eff}(\rr) = -\kB T \frac{v_0 f}{d D_r D_\text{act}} I(\rr).
\end{equation}
In other words, when $f>0$ (positive phototaxis), particles accumulate at the maxima of $I$. On the contrary, when $f<0$ (negative phototaxis), particles accumulate at the minima of $I$: this is the expected result in this limit.

\subsection*{Effective potential}

\subsubsection*{Analysis of the extrema}

We now study the effective potential $U_\text{eff}$ and the stability of its extrema. For simplicity, we assume that the system is effectively one-dimensional, and we compute the derivative of the effective potential:
\begin{equation}
	U'_\text{eff}(x) = - \kB T \frac{v_0 f(I(x))}{dD_r D_\text{act}} \left[1+\frac{v_0 \tau_{\rm m} D }{2(d+2) D_\text{act} D_r} f'(I(x))  I'(x)^2\right]  I'(x),
\end{equation}
and its second derivative:
\begin{align}
	U''_\text{eff}(x) =
	- \kB T \frac{v_0 f(I(x))}{dD_r D_\text{act}} \Bigg\{& f'(I(x))I'(x)^2\left[1+\frac{v_0 \tau_{\rm m} D }{2(d+2) D_\text{act} D_r} f'(I(x))  I'(x)^2\right] \nonumber\\
	&+ \frac{v_0 \tau_{\rm m} D }{2(d+2) D_\text{act} D_r} f(I(x)) I'(x)^2 [I'(x)^2 f''(I(x))+2f'(I(x))I''(x)] \nonumber\\
	&+f(I(x)) I''(x) \left[1+\frac{v_0 \tau_{\rm m} D }{2(d+2) D_\text{act} D_r} f'(I(x))  I'(x)^2\right] \Bigg\}.
\end{align}

The extrema of the effective potential $U_\text{eff}$ are reached at positions such that:
\begin{itemize}
	\item $I'(x)=0$ (i.e. extrema of $I$),
	\item $f(I(x))=0$ (i.e. positions such that $I(x)=I_\text{thr}$),
	\item other special values of $x$ which verify:
	\begin{equation}
		1+\frac{v_0 \tau_{\rm m} D }{2(d+2) D_\text{act} D_r} f'(I(x))  I'(x)^2=0.
	\end{equation}
\end{itemize}

We consider a specific shape for $I(x)$, which is shown on Fig. \ref{threshold_function}b and which is consistent with the shape studied in both experimental and numerical approaches. It has a minimum at $x=0$ and two maxima at $x=\pm x_\text{max}$, where it reaches the value $I_\text{max}$.

We first study the stability of the extrema of $I(x)$, depending on the value of the threshold $I_\text{thr}$:
\begin{itemize}
	\item if $I_\text{thr}< I(0)$: $x=0$ is stable, $x=\pm x_\text{max}$ are unstable,
	\item if $ I(0) < I_\text{thr} < I_\text{max}$: $x=0$ is unstable, $x=\pm x_\text{max}$ are unstable,
	\item if $I_\text{thr} > I_\text{max}$: $x=0$ is unstable, $x=\pm x_\text{max}$ are stable.
\end{itemize}
Importantly, the stability of these points only depend on the intensity threshold, and not on the memory time $\tau_{\rm m}$ of the ABP.

We also study the stability of the points which are such that $I(x)=I_\text{thr}$ , depending on the value of the threshold $I_\text{thr}$. Let $x_\text{thr}$ be such a point,  i.e. $I(x_\text{thr})=I_\text{thr}$. The second derivative of the effective potential evaluated at $x_\text{thr}$ reads:
\begin{equation}
	U''_\text{eff}(x_\text{thr}) = -\frac{\kB T v_0 }{d D_r D_\text{act}} \underbrace{f'(I_\text{thr})}_{<0}\underbrace{I'(x_\text{thr})^2}_{>0} \left[1+\frac{v_0 \tau_{\rm m} D }{2(d+2) D_\text{act} D_r} \underbrace{f'(I_\text{thr})}_{<0}  \underbrace{I'(x_\text{thr})^2}_{>0}\right]
\end{equation}
Therefore, this fixed point becomes unstable when $\tau_{\rm m}$ is greater than a critical value, given by:
\begin{equation}
	\tau_{\rm m}>\tau_c = \frac{2(d+2) D_\text{act} D_r}{v_0 D} \left( \frac{-1}{f'(I_\text{thr}) I'(x_\text{thr})^2} \right).
	\label{tau_c}
\end{equation} 
This is the main result of our analytical theory, which we comment on in the main text.

\section*{Supplementary figures}

\begin{figure}[H]
    \centering
    \includegraphics[width= 0.5\linewidth, page=6]{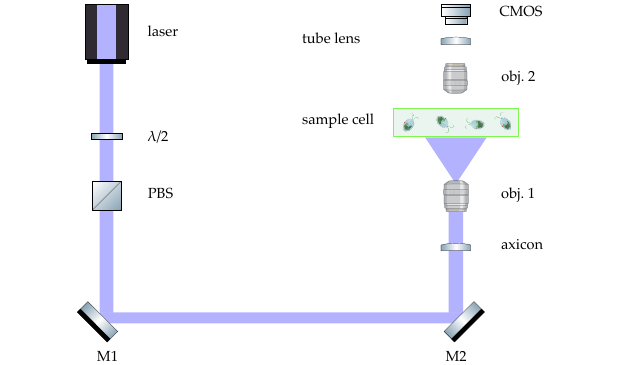}
    \caption{\textbf{Experimental setup.} 
    A laser and an axicon are used to generate a ring-shaped illumination for the microalgae.
    The laser is directed to the sample via two mirrors (M1, M2) and an axicon illuminating the back aperture of an objective (obj. 1). Microalgae are imaged via a second objective (obj. 2), a tube lens and a camera (CMOS). The trajectories are then reconstructed using a custom software. The ring size can be adjusted by changing the beam size and the position of the axicon with respect to the sample. The laser power is controlled by a half-waveplate ($\lambda$/2) and a polarising beamsplitter (PBS).
   }
    \label{figS_exp}
\end{figure}

\begin{figure}[H]
    \centering
    \includegraphics[width= 0.7\linewidth, page=5]{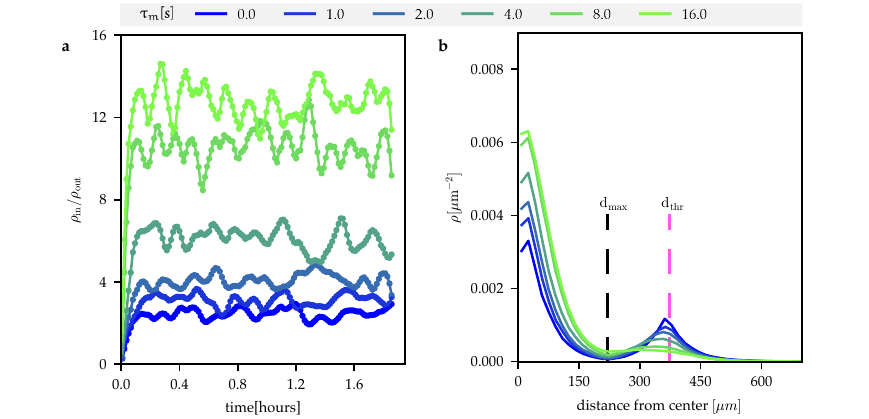}
    \caption{
    \textbf{Accumulation dynamics in a thin ring illumination.} 
    Numerical \rjj{simulation results for} the dynamics of the number of cells (\textbf{a}) and steady-state spatial profiles (\textbf{b}) in a ring illumination with inner radius of \SI{150}{\micro\metre} and outer radius of \SI{250}{\micro\metre}. As in \autoref{fig4}, the simulations show a good agreement with the experimental results (\autoref{fig2}). Compared to the wider ring in the main text, the thinner geometry shown here leads to stronger accumulation at shorter memory times, as cells can more easily reach the inner region via fluctuations. Source data are provided as a Source Data file.
    }
    \label{figS_simSmallRing}
\end{figure}

\begin{figure}[H]
    \centering
    \includegraphics[width= \linewidth, page=7]{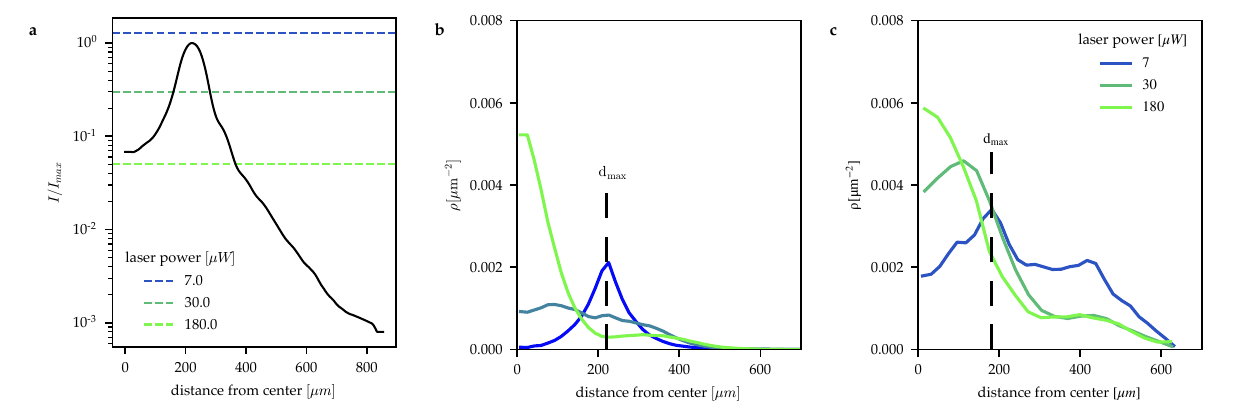}
    \caption{
    \textbf{Radial density profiles under varying illumination intensity.}
   The position at which microalgae switch from positive to negative phototaxis is set by the intensity threshold, whose distance from the centre of the illumination can be tuned by adjusting the laser power. In \textbf{a}, the threshold location is indicated by the intersection of the intensity profile with horizontal lines representing different power levels. Simulated radial density profiles (\textbf{b}) show that \rjj{cells accumulate inside the ring when the threshold value is lower than the well intensity (light green)}, consistent with \autoref{fig2}. As the power decreases and the threshold moves above \rjj{the well intensity (dark green)}, the accumulation shifts outward, \rjj{until cells concentrate near $d_{\rm max}$ when the threshold intensity becomes larger than the peak intensity (blue)}. Experimental measurements (\textbf{c}) show a similar trend, confirming that the shape of the accumulation profile is governed by the relative value of \rjj{the intensity threshold with respect to the peak intensity and well intensity}. All profiles were measured under dilute conditions over a 2 minutes duration, matching the simulation parameters. Source data are provided as a Source Data file.
    }
    \label{figS_intensities}
\end{figure}

\begin{figure}[H]
    \centering
    \includegraphics[width= \linewidth]{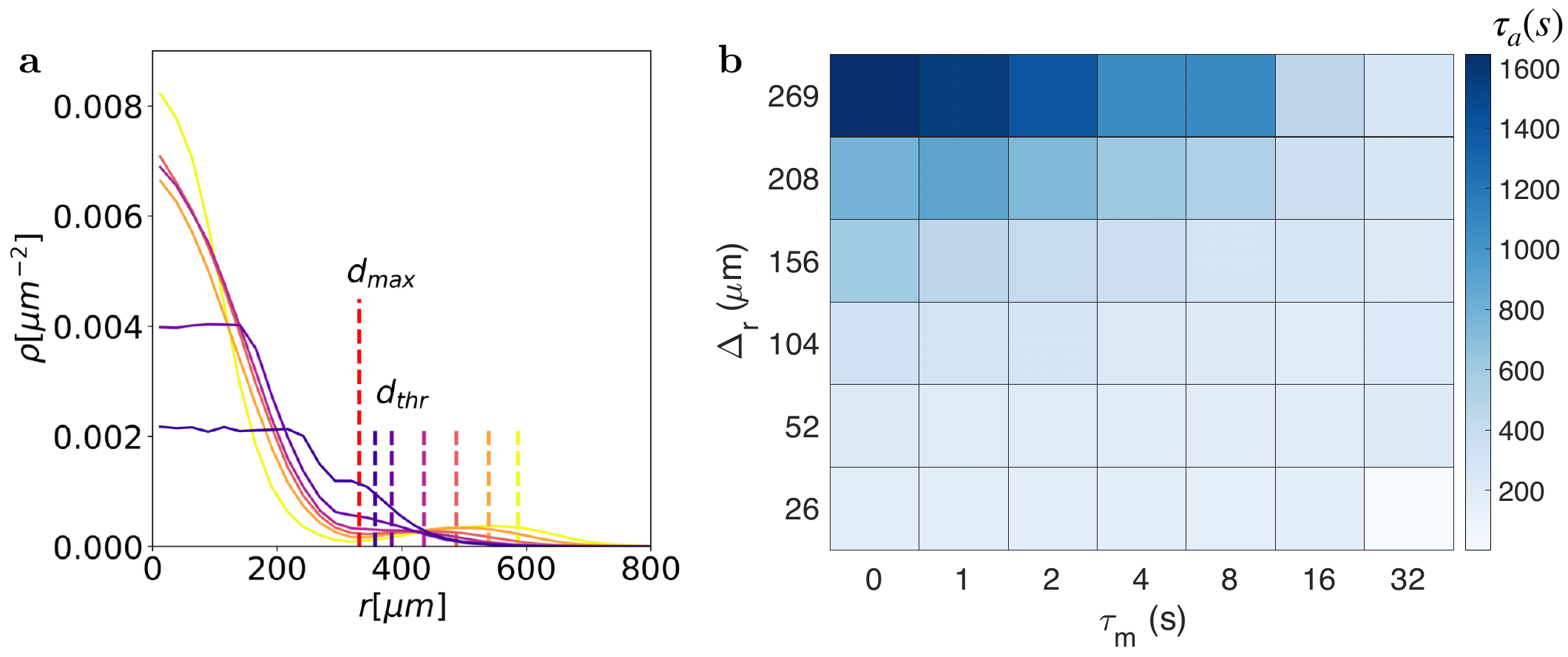}
    \caption{
    \textbf{Sweep in parameter space $(\Delta_r,\tau_m)$.} (\textbf{a}) Steady-state spatial distributions of the agent for $\tau_m=16$s and various $\Delta_r$ (same colorcode as \autoref{fig4}\textcolor{mygreen}{e}). For small thickness $\Delta_r$ (purple curves), a clear shoulder appears at $r=d_{\rm max}$ as the agents tend to spend time diffusing around the intensity maximum before switching to negative phototaxis and following the gradient within the well. (\textbf{b}) Characteristic time $\tau_a$ to reach the steady-state as a function of $\tau_m$ and $\Delta_r$: the wider the ring or the shorter the memory, the longer it takes to reach a steady-state. Source data are provided as a Source Data file.
    }
    \label{figS_optimum}
\end{figure}

\begin{figure}[H]
    \centering
    \includegraphics[width= \linewidth, page=22]{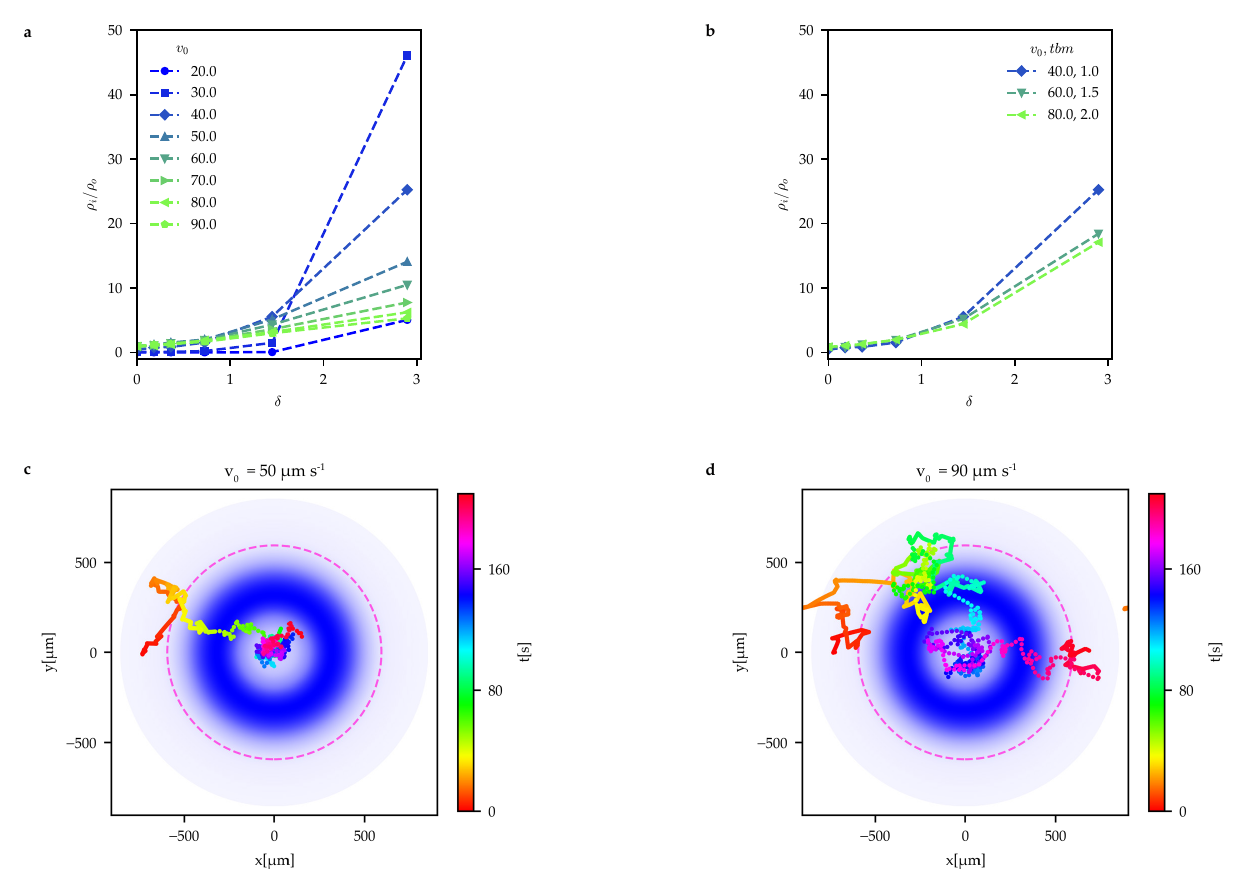}
    \caption{
    \textbf{Dependence of accumulation on self-propulsion speed and tumble rate.
    }
    \textbf{a)} Accumulation ratio $\rho_{\mathrm{in}}/\rho_{\mathrm{out}}$ as a function of the dimensionless parameter $\delta = v_0 \tau_m / (d_{\mathrm{thr}} - d_{\mathrm{max}})$ for increasing swimming speeds $v_0$. Accumulation peaks at intermediate speeds (approximately \SI{30–40}{\micro\metre\per\second}).
    \textbf{b)} When the persistence length is fixed across speeds by adjusting the tumbling rate (tbm), the curves collapse onto a single trend, confirming that accumulation is governed by the effective memory length relative to the ring geometry.
    \textbf{c–d)} Simulated trajectories at $v_0 = 50$ and $90\ \mu\mathrm{m,s^{-1}}$ under the same tumbling rate. Faster (and therefore more persistent) swimmers are  
    \rjj{more likely to traverse the whole width of the ring in a few runs, even in a negative phototactic state, and escape the illuminated region, limiting the trapping efficiency.}
    Trajectories are colour-coded by time; solid and \rjj{dotted} segments indicate \rjj{positive and negative phototactic state of the agent}, respectively. Source data are provided as a Source Data file.
    }
    \label{figS_Delta}
\end{figure}

\begin{figure}[H]
    \centering
    \includegraphics[width= \linewidth, page=9]{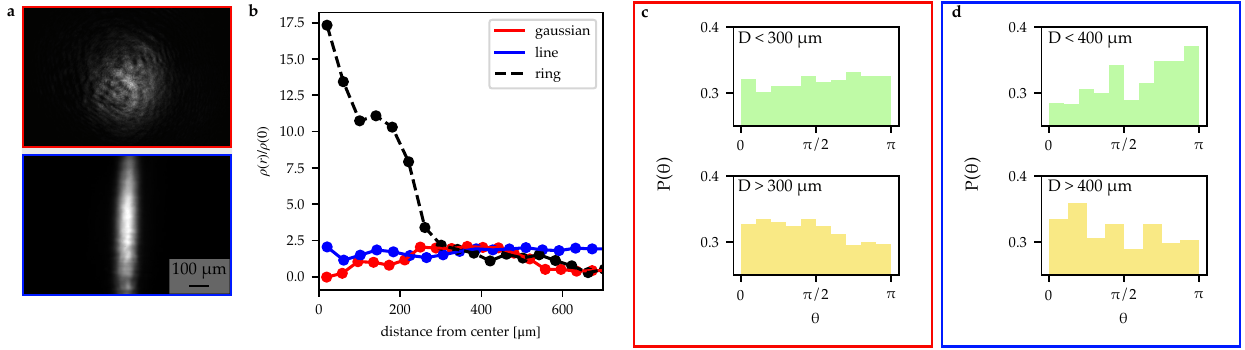}
    \caption{
    \textbf{Phototactic response under different light geometries.}
    \textbf{a)} Experimental images of Gaussian (top) and line (bottom) illumination patterns.  
    \textbf{b)} Normalised radial density profiles $\rho(t)/\rho(0)$ measured after 2 minutes of exposure to ring (black), Gaussian (red), and line (blue) illumination. The line profile remains flat, indicating no accumulation, while the Gaussian profile presents a broad and moderate peak around $300-400 {\rm \mu m}$. Here, expected accumulation at $I_{\rm thr}$ is probably impeded by:
    i) the memory effects leading to cells oscillating around $d_{\rm thr}$ with large excursions (therefore flattening the expected accumulation), and ii) a distribution of intensity thresholds among the algal population.
    \textbf{c–d)} Angular distributions of swimming direction $P(\theta)$ in the Gaussian c) and line d) geometries, computed separately for cells located near the \rjj{centre (top) and farther away (bottom).
    In both cases, cells tend to swim along the light gradient when far from the centre, and reverse direction when closer to it—consistent with a switch from positive to negative phototaxis at some distance to the peak intensity ($\sim$\SI{300}{\micro\metre} for the Gaussian case, $\sim$\SI{400}{\micro\metre} for the line).}
    All patterns were calibrated to yield the same energy density over the illuminated region. Source data are provided as a Source Data file.
    }
    \label{figS_lineGau}
\end{figure}

\begin{figure}[H]
    \centering
    \includegraphics[width= \linewidth, page=10]{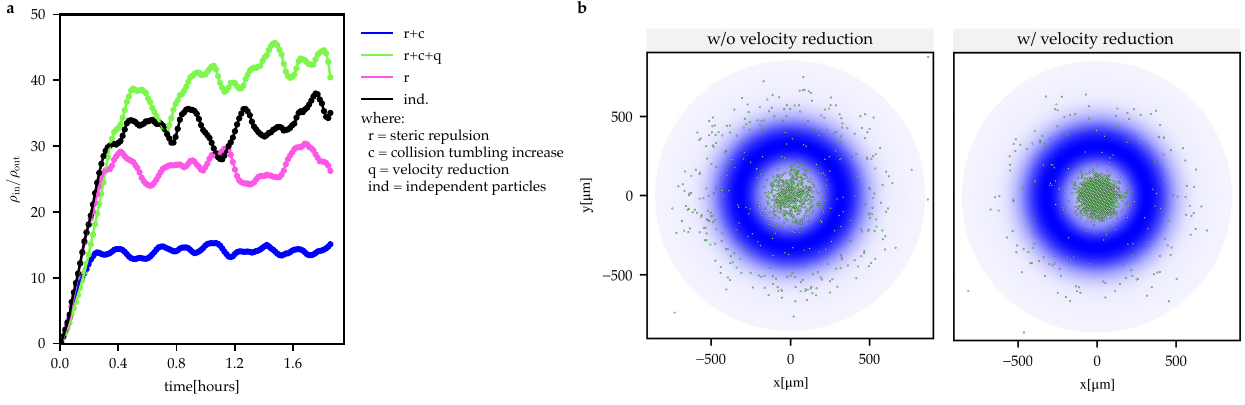}
    \caption{
    \textbf{Role of density-dependent interactions in accumulation dynamics.}
    \textbf{a)} Time evolution of the density ratio between the inside and outside of the ring, $\rho_\text{in} / \rho_\text{out}$, computed from simulations including different interaction terms. Accumulation persists in the absence of collective effects (black: independent particles), confirming that phototactic memory alone is sufficient to drive localisation. Adding steric repulsion (magenta), collision-induced tumbling (blue), and velocity reduction at high density (green) progressively alters the dynamics. In particular, motility quenching enhances accumulation and leads to more ordered spatial organisation inside the ring.
    \textbf{b)} Snapshots from simulations with (right) and without (left) motility quenching show that velocity reduction leads to more ordered cell distributions. Simulations use a ring geometry with inner and outer radii of \SI{150}{\micro\metre} and \SI{420}{\micro\metre}, respectively. Source data are provided as a Source Data file.
    }
    \label{figS_Interaction}
\end{figure}

\begin{figure}[H]
    \centering
    \includegraphics[width= \linewidth, page=23]{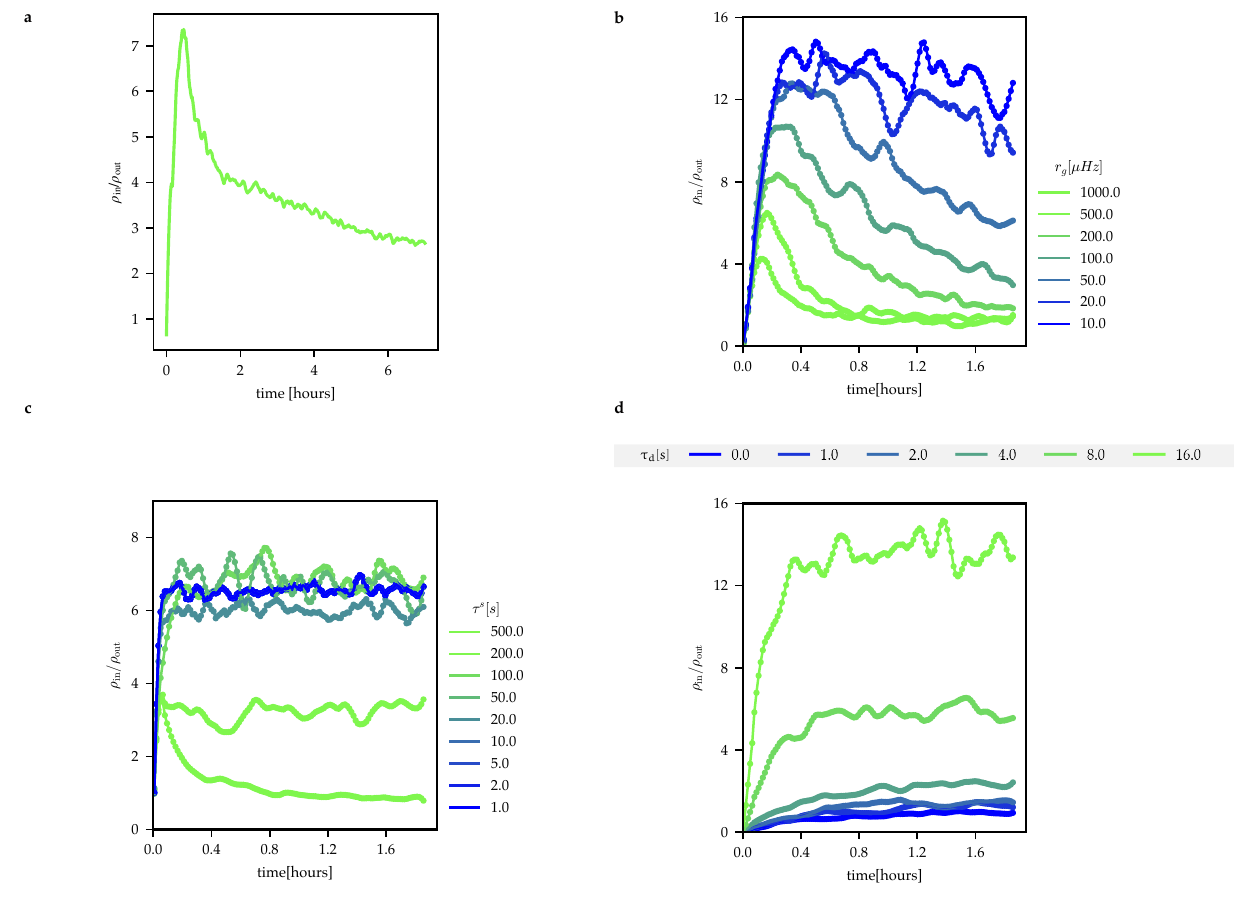}
    \caption{
    \textbf{Phototactic adaptation reduces long-time accumulation.}
    \textbf{a)} Experimental time evolution of the accumulation ratio $\rho_{\mathrm{in}}/\rho_{\mathrm{out}}$ under constant ring illumination shows that accumulation peaks at around one hour and then gradually declines, \rjj{interpreted as}  a loss of phototactic sensitivity at long times.
    \textbf{b)} Numerical simulations of a minimal adaptation model, where the phototactic response decays over time with a rate $r_g$. Faster decay leads to earlier loss of accumulation, reproducing the experimental trend in a).
    \textbf{c)} Simulations based on the intracellular-species model \rjj{(S model) with different deactivation times $\tau^s$ either show steady-state accumulation (small $\tau^s$) or time-decaying accumulation (large $\tau^s$), when the parameter $\gamma^s$ is fixed to $5 \times 10^{-5}\,\mathrm{s^{-1}}$.} 
    \textbf{d)} An alternative mechanism \rjj{for accumulation} is based on a fixed delay $\tau_d$ between light sensing and \rjj{phototactic switch. This delayed response alone generates memory-like accumulation, reaching levels comparable to the explicit memory model. Source data are provided as a Source Data file.}
    }
    \label{figS_Adaptation}
\end{figure}

\begin{figure}[H]
    \centering
    \includegraphics[width= 0.5\linewidth, page=17]{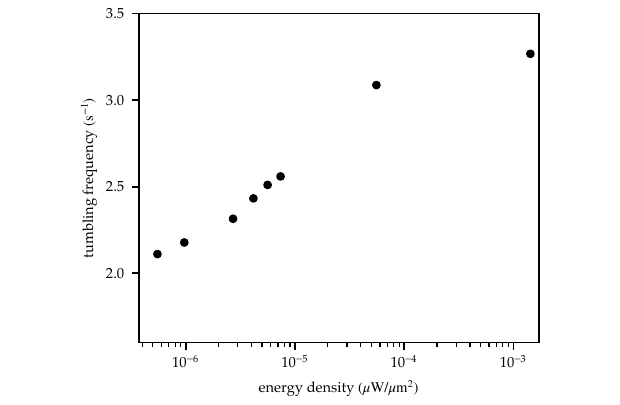}
    \caption{
    \textbf{Light intensity modulates the tumbling frequency.}
    Tumbling frequency measured as a function of light intensity under spatially uniform illumination generated by a LED panel.
    The observed increase confirms that higher light levels induce more frequent reorientations, providing the basis for constructing a spatially dependent tumbling rate field $f_R(I(\mathbf{r}))$ in the simulations. This function defines how the local light intensity modulates cell reorientation, forming the effective phototactic response field \rjj{$I_{\rm tumb}(\mathbf{r})=f_R(I(\mathbf{r}))$} used in the model. While such modulation is present in experiments, it plays a secondary role: on its own, it tends to concentrate cells at the intensity maximum, and cannot account for the ring-shaped accumulation, which is instead driven by memory. Source data are provided as a Source Data file.
    }
    \label{figS_TumbMaster}
\end{figure}

\begin{figure}[H]
    \centering
    \includegraphics[width= 0.5\linewidth, page=20]{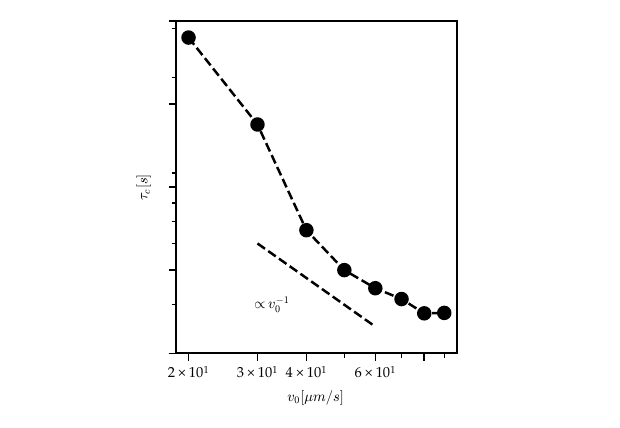}
    \caption{
    \textbf{Critical memory time for accumulation as a function of swimming speed.}  
    Critical value of the memory time $\tau_c$ extracted from simulations, above which accumulation at the centre of the ring is observed, as a function of the swimming speed $v_0$. The data confirm that $\tau_c$ decreases with increasing $v_0$, consistent with the analytical prediction $\tau_c \propto 1/v_0$ (dashed line). The threshold $\tau_c$ is defined as the value at which the density ratio $\rho_{\mathrm{in}} / \rho_{\mathrm{out}} \geq 2$ \rjj{in the steady-state}. Source data are provided as a Source Data file.
    }
    \label{figS_SimuVsAnalytical}
\end{figure}

\begin{figure}
	\begin{center}
		\includegraphics[width= 0.75\linewidth, page=21]{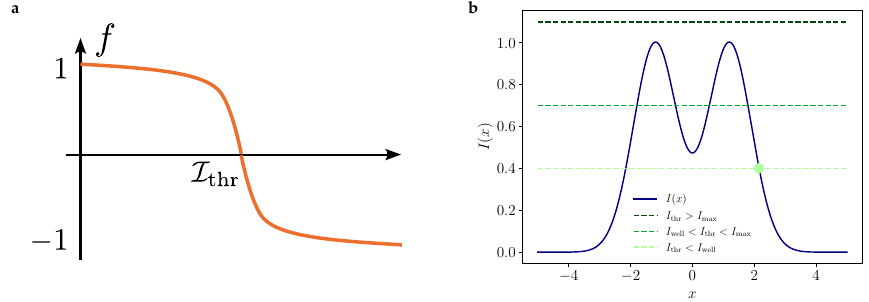}
	\end{center}
	\caption{Threshold-based response function and intensity landscape.
		\textbf{a)} Sketch of the threshold function $f(I)$ used to determine the sign of the phototactic response, switching from positive to negative at the intensity threshold $I_T$.
		\textbf{b)} Example intensity profile $I(x)$ with three illustrative cases for the threshold $I_T$: above the maximum ($I_T > I_{\mathrm{max}}$), between the centre and maximum ($I(0) < I_T < I_{\mathrm{max}}$), and below the central intensity ($I_T < I(0)$). These scenarios determine whether and where cells reverse their phototactic response in the light field.}
	\label{threshold_function}
\end{figure}

\newpage
\putbib  

\end{bibunit}

\end{document}